\documentclass[12pt]{article}
\usepackage[T1]{fontenc}
\usepackage[utf8]{inputenc}
\usepackage{lmodern}
\usepackage{amsmath,amssymb,amsfonts}
\usepackage{booktabs}
\usepackage{tabularx}
\usepackage{graphicx}
\usepackage{todonotes}
\usepackage[english]{babel}
\usepackage{hyperref}
\usepackage[caption=false]{subfig}

\newcommand{\re}{\mathbb R}

\usepackage{placeins}

\let\Oldsection\section
\renewcommand{\section}{\FloatBarrier\Oldsection}

\let\Oldsubsection\subsection
\renewcommand{\subsection}{\FloatBarrier\Oldsubsection}

\let\Oldsubsubsection\subsubsection
\renewcommand{\subsubsection}{\FloatBarrier\Oldsubsubsection}

\graphicspath{{Figures/}}
\newcommand{\Cs}{\mathcal{C}}
\newcommand{\p}{\preceq}

\newcommand{\av}[1]{\langle #1 \rangle}

\newcommand{\twod}{\mathrm{2d}}

\newcommand{\mink}{{\mathbb M}^2}

\newcommand{\bb}{\bar\beta}
\newcommand{\bbS}{\bar S}
\newcommand{\bbF}{\bar F}
\newcommand{\bC}{\bar C}

\newcommand{\fut}{\mathrm{Fut}}
\newcommand{\past}{\mathrm{Past}}

\usepackage{fullpage}

\begin{document}

\title{Finite Size Scaling  in  \\  $\twod$ Causal Set Quantum Gravity
}
\date{\today}
\author{Lisa Glaser${}^a$, Denjoe O'Connor${}^b$ and Sumati Surya${}^c$\\
${}^a$Radboud University, Njimegen, \\  ${}^b$DIAS, 10 Burlington Road
Dublin 4
Ireland \\
${}^c$ Raman Research Institute, CV Raman Ave,
Bangalore, India, \\
and Perimeter Institute, Waterloo, Canada }
\maketitle
\begin{abstract}
  We study the $N$-dependent behaviour of $\twod$ causal set quantum
  gravity. This theory is known to exhibit a phase transition as the
  analytic continuation parameter $\beta$, akin to an inverse
  temperature, is varied. Using a scaling analysis we find that the
  asymptotic regime is reached at relatively small values of $N$.
  Focussing on the $\twod$ causal set action $S$, we find
  that $\beta \av{S} $ scales like $ N^\nu$ where the scaling exponent $\nu$
  takes different values on either side of the phase transition. For
  $\beta > \beta_c$ we find that  $\nu=2$ which is consistent with our
  analytic predictions for a
  non-continuum phase in the large $\beta$ regime. For $\beta<\beta_c$
  we find that $\nu=0$, consistent with a
  continuum phase of constant negative curvature thus suggesting a
  dynamically generated cosmological constant.  Moreover, we find
  strong evidence that the phase transition is first order.  Our results
  strongly suggest that the asymptotic regime is reached in $\twod$
  causal set quantum gravity for {$N \gtrsim 65$}.
  \end{abstract}

\section{Introduction}


In causal set theory (CST) space-time is regarded as fundamentally
discrete~\cite{cst}.  Approaching quantum gravity from this vantage
point makes it possible to use computer simulations to explore
non-perturbative features of the
theory~\cite{2dqg,hh,henson_onset_2015}.  The system size that can be
examined in computer simulations is always limited, and particularly
so in CST, where non-locality leads to greater complexity.  A finite
size scaling analysis is therefore important in order to ensure that
the results obtained for a given value of $N$ can be generalized to
larger $N$.  Only if we find a convergence with $N$ is it possible to
generalize from finite values to the asymptotic regime.



Scaling with $N$ is an important question for all lattice-type numerical
simulations, whether in lattice gauge theory or quantum gravity, where
large $N$ convergence means that crossover transients due to finite
size effects are suppressed.  A standard way to assess this is to vary
$N$ and look for consistent scaling behaviour of relevant observables
in order to extract scale invariant quantities. The physically
reasonable assumption is that consistent scaling is evidence for
asymptotic behaviour.

Such questions are of particular relevance to CST where numerical
simulations of the dynamics are carried out for relatively small
system size $N$\cite{cst,2dqg,hh, henson_onset_2015}. CST assumes a
discrete or atomistic nature of spacetime as its fundamental kinematic
hypothesis \cite{cst}.  It is founded on the idea that discreteness
and causality encode approximate Lorentzian geometry.  Discreteness is
implemented by replacing the spacetime continuum by locally finite
partially ordered sets, or causal sets, with the order relation
corresponding to the causal relation ~\cite{cst}. The condition of
local finiteness captures the main hypothesis of the theory, namely
that there is a {\it fundamental} spacetime discreteness or atomicity,
so that every spacetime region of finite volume contains a finite
number of causal set elements.

Causal sets that are approximated by the continuum are ``random
lattices'' generated via a  Poisson process. While this implies
Lorentz invariance ~\cite{bhs}, it  comes at the price of locality  since the resulting graph is not of
fixed or even finite valency. For
example, a causal set that is approximated by Minkowski spacetime is
not a finite valency graph; the combination of discreteness with
Lorentz invariance ensures that every element in the causal set has
infinite nearest neighbours. Non-locality is therefore  a key feature of
continuum-like causal sets and provides  a promising avenue  for
phenomenological exploration~\cite{belenchia_low_2016,saravani_off-shell_2017}.

A major advance in the causal set program has been the development of a
causal set version of the Einstein-Hilbert action --- the
Benincasa-Dowker (BD) action $S$, which incorporates this non-locality.
The quantum partition function or sum-over-histories over the space of
finite element causal sets can thus be constructed, with each causal
set weighted by the quantum measure $\exp({i S/\hbar})$. This defines a
theory of quantum gravity \cite{2dqg}. 

By introducing a parameter $\beta$ akin to the inverse
temperature this  quantum partition function can be rendered into a
Gibbs ensemble of finite element causal sets weighted by the BD action
\cite{sorkin_is_2009,2dqg}. This makes the theory amenable to
numerical analysis and has yielded interesting results \cite{2dqg,hh, henson_onset_2015}.
However, computational constraints arise when working with non-local
graphs of
 large connectivity, and put practical limits on the size
$N$ of
 causal sets that can be used in numerical simulations.  When simulating the quantum dynamics of causal sets
using Markov Chain
 Monte Carlo (MCMC) algorithms, the non-locality
increases the
 complexity drastically and slows down thermalisation
times. The
 explorations of the parameter space to understand the
asymptotic regime require a large number of independent simulations
and are therefore resource demanding.  While efforts are underway
 to
improve algorithms, the fundamental limitation coming from the
complexity of the graphs cannot be easily overcome.

 In light of
this it becomes all the more important to understand the
 ``finite
size effects'' in CST by examining the $N$ dependence of
 various
physically relevant quantities\footnote{Since discreteness is fundamental in
CST, however, the attitude
 to finite
size effects differs from theories in which discretisation is used as  an ultraviolet regulator. Rather, the
situation is similar to condensed matter physics where a finite
value of $N$, however small, is {\it{not}} unphysical.}.
Extrapolating to large $N$ helps in acertaining macroscopic properties of the
asymptotic
 regime, which can be relevant for the continuum
approximation as well
 as for phenomenology.

Using MCMC methods
it was shown in~\cite{2dqg} that 2d CST exhibits
 two distinct phases
as one varies over $\beta \in \re^+$. The high
 temperature phase is
a continuum-like phase which we refer to as the
 $\Pi_{+}$ phase and
the other is a crystalline or non-continuum like
 phase, we label
$\Pi_{-}$ which occurs at low temperatures. The continuum like
behaviour is consistent with the infinite temperature limit in which
the system is known to be dominated by causal sets that are
approximated by 2d Minkowski spacetime~\cite{ezs,winkler}. The low
temperature behaviour on the other hand is a new phase characterised
by causal sets with very high graph connectivity, which has
physically
 interesting consequences for the 2d Hartle-Hawking wave
function, as
 shown in~\cite{hh}.  

The results of~\cite{2dqg} were obtained for causal sets with a few
values of $N$ and $\epsilon$, all of which showed the same qualitative
features. However, no scaling analysis was done, and it was unclear if
the phase transition would survive the large $N$ limit.  Indeed, it
is the purpose of this paper to fill in this gap by doing a finite
size scaling analysis and obtain convergence. In the process we are able to extract interesting
physical information about the detailed nature of the phases and find
strong evidence that the transition is first order.

Our principal results are
\begin{itemize}
\item For a fixed non-locality parameter $\epsilon$, the  phase
  transition is shown to occur at a critical value $\bb_c$  of the
  rescaled inverse temperature $\bb=\beta N$,
 where $\bb_c$  varies with $\epsilon$ as   $\bb_c \sim \frac{1}{\epsilon^2}$
 in the large $N$ limit.
\item We find that for high, but finite, temperature the continuum
  phase is characterised by $\av{S} \propto N$ and   corresponds to
  a
  constant negative curvature in the continuum approximation.
We interpret this phase as an emergent anti-de
  Sitter space, which at infinite temperature becomes Minkowski
  space.
\item In the small temperature phase we find that $\av{S} \propto N^2$,
  which is consistent with our analysis that the zero temperature
  limit should be dominated by non-manifoldlike bilayered posets.
\item We estabilish that the two phases are separated by a {\it first
    order}
  phase transition.
\item The system enters the asymptotic regime for $\twod$ CST  for $N\sim 65$.
\end{itemize}

Our paper is organised as follows. In Section \ref{two.sec} we
describe $\twod$ CST in some detail and review the results of
\cite{2dqg}. In Section \ref{three.sec} we define the large $N$
asymptotic scaling exponents for $\av{S}$ on either side of the phase
transition. We give analytic arguments for what the exponent should be
in the limit of large $\beta$, based on the analysis of the
action.
In Section \ref{four.sec} we present the main results
of this paper. We have generated an extensive data set using MCMC
simulations for $N$ ranging from $20$ to $90$ and for $\epsilon$
ranging from $0.1$ to $0.5$, as shown in Table \ref{simdata.tab}.
This allows us to perform a detailed study and obtain not only the
large $N$ scaling of $\av{S}$ before and after the transition, but also
an estimate for the function $\av{S}$ itself. We first show that
the estimated critical inverse temperature $\beta_c$ scales like ${N}^{-1}$.
Using the rescaled inverse temperature $\bb = \beta N$, we
find that for the partition function $Z$,  $\ln Z=-\beta F$ scales as
$-\bb\bar F N^{\nu_\pm}$ (Eqn \ref{scalingF})
with
the scaling exponents $\nu_+=1$, and $\nu_-=0$ on either side of  the phase
transition. Hence we  find that $\ln Z$ is scale invariant before the phase
transition and is extensive, i.e., scales with $N$, after the phase
transition. Finally we demonstrate that the transition is of first
order. In Section \ref{five.sec} we discuss some open questions.

\section{A review of 2d causal set theory}
\label{two.sec}
\subsection{Mathematical Preliminaries}

We remind  the reader of some key definitions in CST and
refer them to the reviews~\cite{sorkin_space-time_1990,surya_directions_2011,sorkin_causal_2003} for more details.
A {\sl causal set}  $C$ is a locally finite partially ordered
set. Thus $C$ is a set with a relation $\p$  which is (i)
reflexive: $x \p x$, (ii) transitive: $x \p y$ and $y \p z$ implies
$x \p z$  (iii) acyclic: $x,y \in \Cs$ and $x \p y \p x$  $\,
\Rightarrow \, x =y$ (iv)
locally finite: $| \fut(x) \cap \past(y)| < \infty $, where  $\fut(x)\equiv\{ z|
x \prec z\}$ and $\past(x)\equiv \{z| z \prec x \}$.

The causal set hypothesis  assumes that continuum spacetime is replaced by a discrete substratum,
the causal set. A causal set $C$ is said to have a {\sl
  continuum approximation} to a spacetime $(M,g)$ if $C$ can be
obtained from a Poisson sprinkling into $(M,g)$ where
$P_v(n)=\frac{1}{n!} (\rho V)^n{\rm e}^{-\rho
  V}$ is the probability of sprinkling $n$ elements into a spacetime
volume $V$.

 Importantly, the assumption of a fundamental   discreteness  implies
 that  the continuum limit itself is unphysical;  it is only the
 continuum approximation that is physically relevant.  Since
 discreteness is not used as a regulator for the continuum,  finite
 size causal sets are not unphysical. The large $N$ or asymptotic
 regime is phenomenologically interesting but not equivalent to the
 continuum regime.

We now write down some useful definitions. $x, y \in C $ are said to
be {\it linked} if $x \prec y$ and $\nexists \, \, z \in C$
such that $x \prec z \prec y$. A link is therefore an irreducible relation
which cannot be obtained by transitive closure.  A set $A$ with no
relations between the elements is called an {\sl antichain} while a
set $B\equiv \{ e_1, \ldots,  e_N \}$  which is totally ordered, $ e_1
\prec e_2 \prec \dots \prec  e_N$ is called a {\sl chain}.

An {\sl  interval} $I(x,y) \equiv \{z| x \prec z \prec y \}$  is said
to be an {\sl $n$-element interval}  if $|I(x,y)|=n$. Thus if $x \prec
y$ is a link, then $I(x,y)$ is a zero interval.
The {\sl abundance} $N_n$ of $n$-element intervals in $C$ is the number of
such intervals  in $C$. $N_n$ is therefore an important covariant, or
label independent,  observable in $C$.


\subsection{2d CST}
The $\twod$  BD action is
\begin{equation}
\label{eq:action2d}
S(C,\epsilon)= 4 \epsilon (N - 2 \epsilon \sum_{n=0}^{N-2} N_n f(n,\epsilon))\;.
\end{equation}
where $N_n$ is the number of $n$ element intervals and
\begin{equation}
\label{fne}
 f(n,\epsilon)=(1-\epsilon)^n\biggl(1-\frac{2\epsilon n}{(1-\epsilon)}+\frac{\epsilon^2n(n-1)}{2 (1-\epsilon)^2}\biggr)\, .
\end{equation}
This action depends on a non-locality scale $l>l_p$, the Planck scale,
and where $\epsilon=l_p^2/l^2 \in (0,1]$.
This introduces $\epsilon$ as a new free
  parameter or ``coupling'' that
  suppresses the fluctuations in the BD action in the continuum
  approximation.

In~\cite{2dcst,2dqg} a two dimensional restriction of CST was defined
by limiting the set of all causal sets  to the set $\Omega(N)$, of $N$
element ``{$\twod$-orders}''.  These are causal sets that can be
embedded into a causal diamond or Alexandrov interval in $\mink$,
via
an order preserving map. Since the map need not be obtained from a
Poisson sprinkling, not all causal sets in $\Omega(N)$ have a
continuum approximation.

Let us define our configurations more precisely.
Beginning with a ``base set'' $S = (1, . . . , N )$ let
$U = (u_1 , u_2 , . . . ,u_N )$ and $ V = (v_1 , v_2 , . . ., v_N )$,
such that $u_i,v_i\in S$, with $u_i=u_j \Rightarrow i=j$, and
$v_i=v_j\Rightarrow i=j$.  $U$ and $V$ are therefore ``totally
ordered'' by the integers, since every element is related to every
other element.  The {\sl $\twod$ order} $C \equiv U \cap V$, with
$e_i\equiv (u_i,v_i)\in C $ where $e_i \prec e_j$ in $C$ iff
$u_i < u_j$ and $v_i < v_j$ ~\cite{ezs,winkler, 2dcst}.

Every $\twod$ order embeds via an order preserving map into $\mink$
since the $U,V$ orders provide a set of lightcone coordinates
$(u_i,v_i)$ for each element $e_i$. Conversely, every causal set
obtained from a Poisson sprinkling into a topologically trivial
interval in a $\twod$ spacetime is a $\twod$ order. Hence the sample space of
$\twod$ orders $\Omega(N)$ includes all causal sets approximated by topologically
trivial $\twod$ spacetime regions as well as those that have no
continuum approximation. While the unrestricted space of $N$ element
causal sets is known to be dominated by non-manifold like causal sets
\cite{kr,henson_onset_2015} the restriction to $\twod$ orders $\Omega(N)$ was shown in
\cite{ezs,winkler} to be dominated by the so-called ``random''
$\twod$ orders, which are approximated by an interval in $\mink$
\cite{2dcst}. This makes the two dimensional restriction
a good testing ground for non-perturbative causal set quantum gravity.

The quantum partition function is
\begin{equation}
\label{qpf}
Z(N) \equiv \sum_{C \in \Omega} {\rm e}^{\frac{i}{\hbar}S(C,\epsilon)}
\end{equation}
where $S(C,\epsilon)$ is the $\twod$ BD action (Eqn \ref{eq:action2d}) with non-locality
parameter $\epsilon$.  Introducing an analytic continuation parameter
$\beta$ akin to the inverse temperature,  we define  a complex parameter
family of partition functions
\begin{equation}
\label{spf}
Z({\beta},N)\equiv\sum_{C\in \Omega}
{\rm e}^{-\frac{\beta}{\hbar}S(C,\epsilon)}.
\end{equation}
Here $\beta=-i$ is the quantum partition function while $\beta \in
\re^+$ is a statistical partition function (and a Gibbs ensemble of
causal sets), which is amenable to numerical analysis. Note that  the
transition from the quantum to the statistical occurs  without
changing the sample space of causal sets, and hence the intrinsically
Lorentzian character of configurations. This marks an important
contrast with Euclidean methods and the usual Wick rotation procedure
which renders the signature
Euclidean, thus removing all trace of causality.


The equilibrium expectation value for any observable $O$  in the
Gibbs ensemble (Equation \ref{qpf}) is given by
\begin{equation}
\av{O}_\beta = \frac{1}{Z}\sum_{C\in \Omega} O(C) \,  {\rm e}^{-\frac{\beta}{\hbar}S(C,\epsilon)}
\end{equation}
and can be numerically obtained using Markov Chain Monte Carlo methods. In
\cite{2dqg} the Markov Chain was generated via the following  exchange move.
Starting with any $\twod$ order $C=U \cap V$ with
$U = (u_1 , u_2 , . . . ,u_N ), V = (v_1 , v_2 , . . ., v_N )$, a
pair of distinct elements $(i,j)\in S\times S$, $i \neq j$  are picked at
random. Then picking $U$ or $V$ again at
random,  either
$u_i$ and $u_j$ or $v_i$ and $v_j$ are exchanged (doing both will give back
the original configuration). For example,  starting  with the
4-element chain $U=(1,2,3,4), V=(1,2,3,4)$, if  $i=2, j=3$ are picked
in $U$, then the new $\twod$ order is $U=(1,3,2,4), V=(1,2,3,4)$, for
which the $e_2, e_3$ elements are now spacelike, but to the past of
$e_4$ and the future of $e_1$  forming  a ``diamond'' causal set. This move
satisfies detailed balance and is ergodic, as demonstrated explicitly
in~\cite{2dqg}.

\begin{figure}
\centering{
\includegraphics[width=6cm,angle =45]{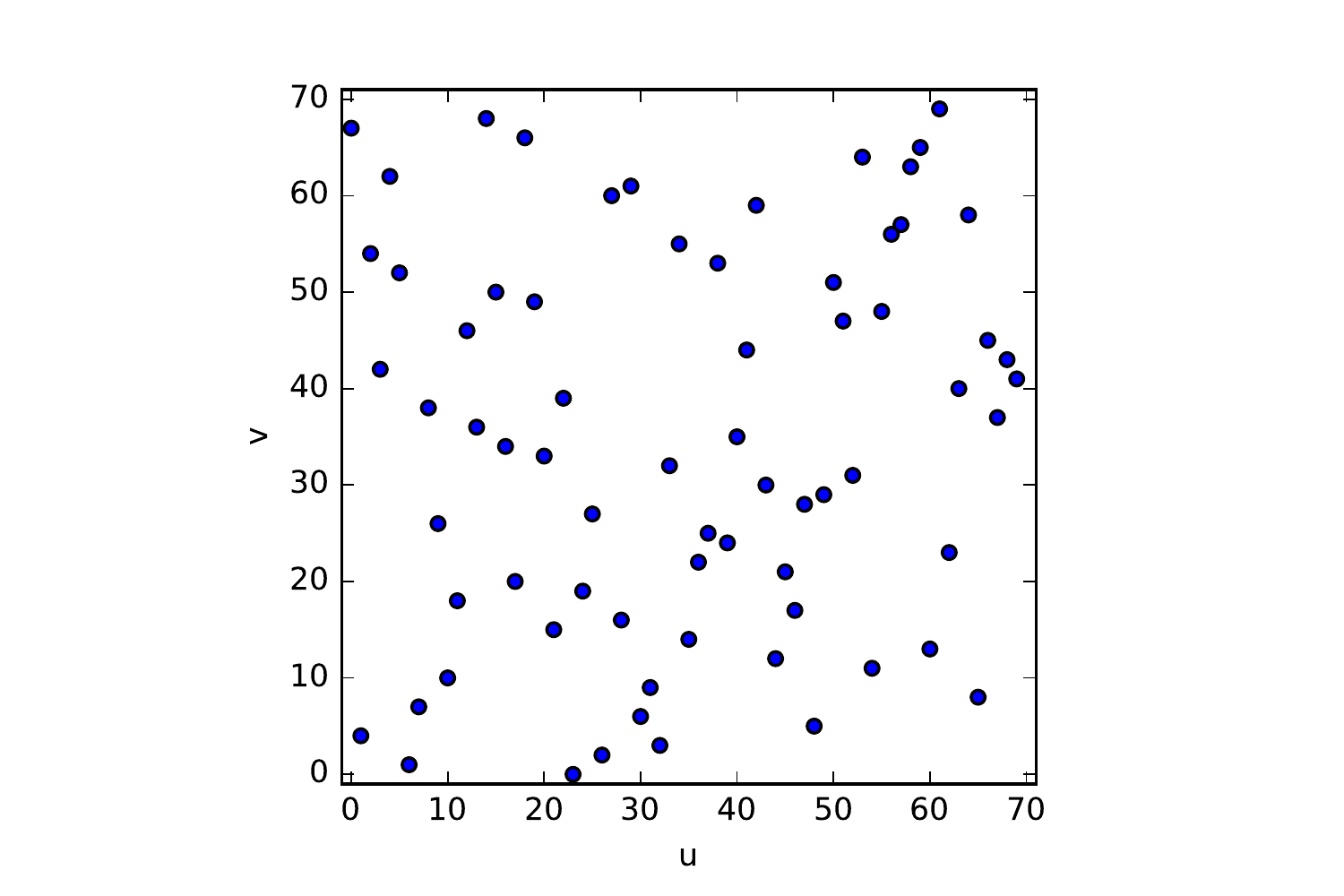}
\includegraphics[width=6cm,angle =45]{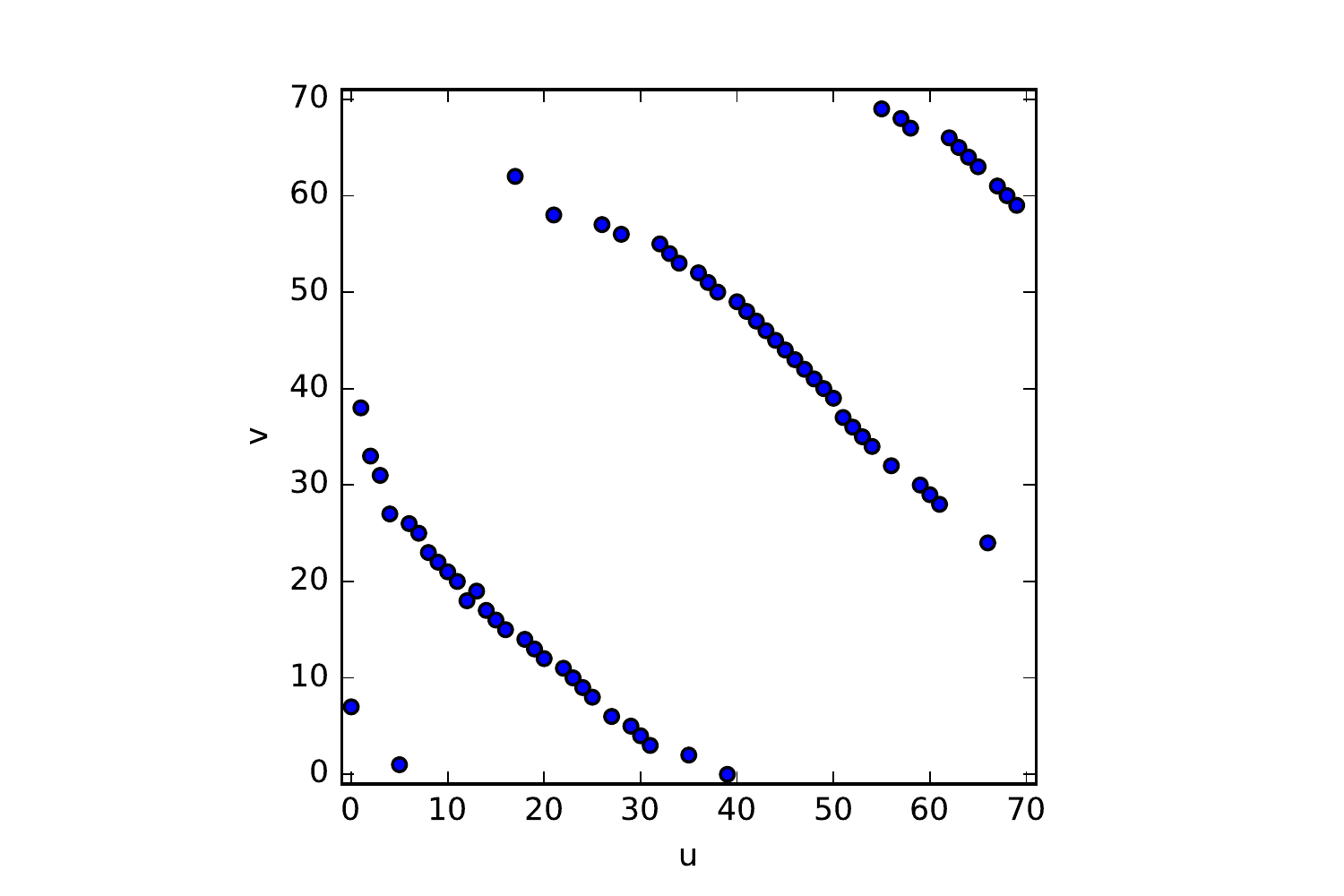}
  \caption{Examples of  causal sets before ($\beta=0.008$) and after ($\beta=0.751$) the phase transition for $N=70$ $\epsilon=0.2$. }
  \label{fig:phases}}
\end{figure}

The expectation values of several covariant observables were calculated
in~\cite{2dqg}, including the action, the abundance of intervals $N_n$,
the number of maximal and minimal elements as well as the ordering
fraction, whose inverse gives the Myrheim-Meyer estimate of the
continuum spacetime dimension. 
$\av{O}_{\beta=0}$ for all these observables gave values consistent with the random $\twod$
orders that dominate in the asymptotic limit~\cite{ezs,winkler}, which are
approximated by $\twod$ Minkowski spacetime. As $\beta$ is increased,
this continuum-like phase gives way,  after a  critical value
$\beta_c$,
to a non-manifold like phase characterised by  ``layered''   causal sets shown in Figure \ref{fig:phases}. The
phases are strikingly similar to what one might expect in an Ising
model: at low temperatures ($\beta> \beta_c$) the dominant phase is
very regular, or crystalline,  which resembles the  ordered phase, while at high temperatures ($\beta<\beta_c$)
it  is random, which resembles the disordered phase.

\begin{figure}
\begin{center} 
\includegraphics[width=0.9\textwidth]{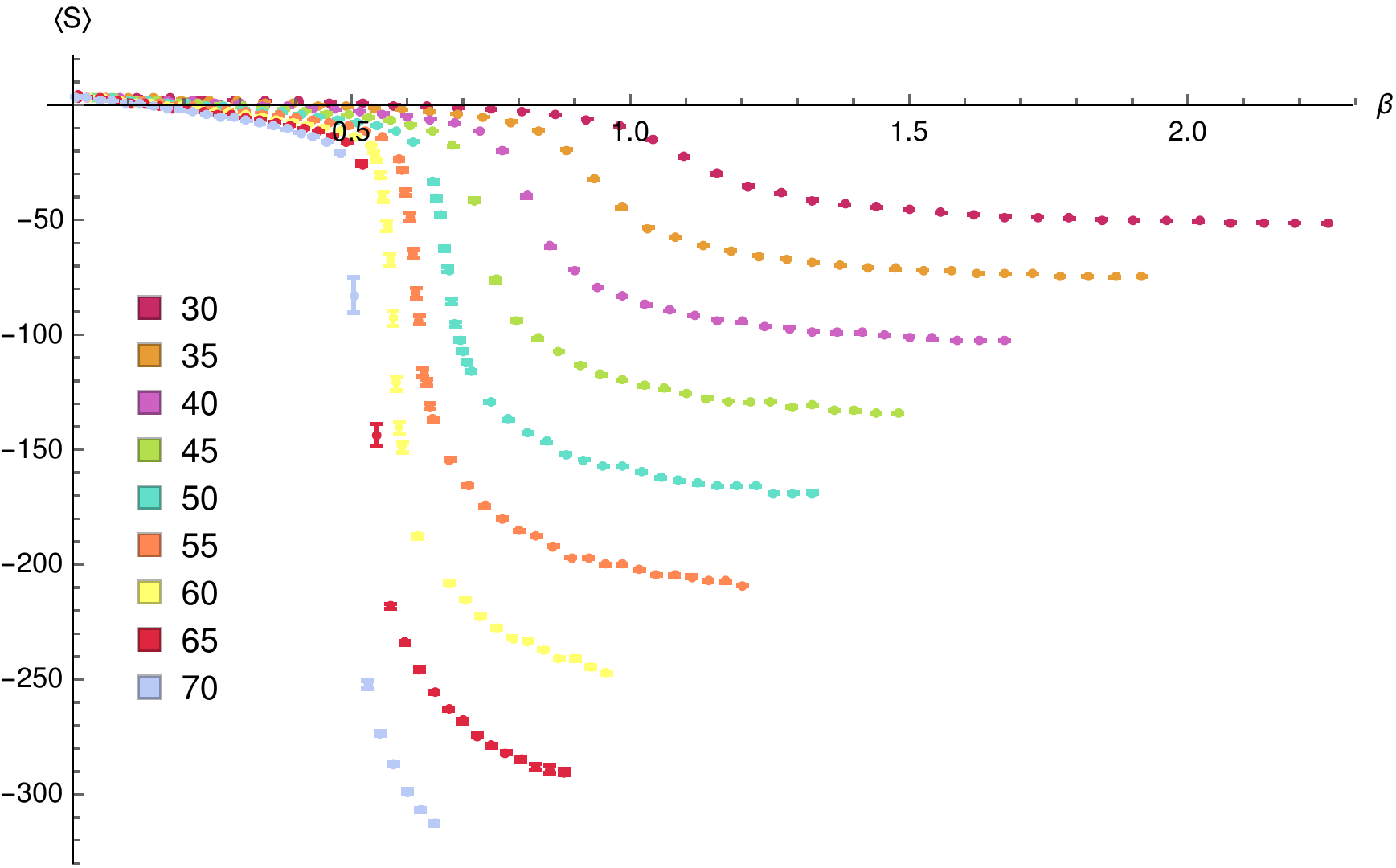}\vskip 1cm
\includegraphics[width=0.9\textwidth]{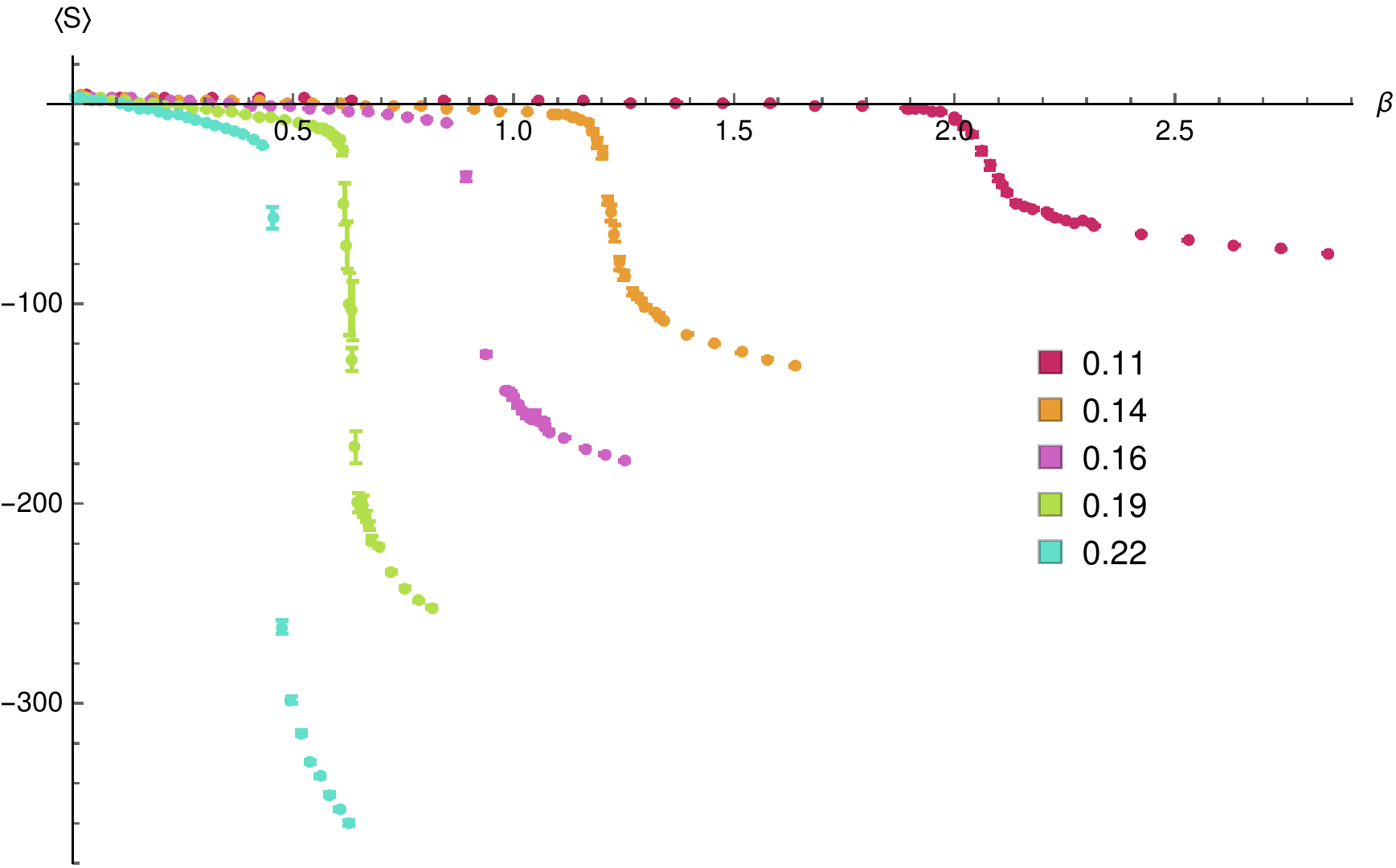}
\caption{\label{unscaled.fig}
 The first figure shows $\av{S}$ v/s
  $\beta$ for various   $N$ values,  for fixed {$\epsilon=0.21$}.
  The second figure  shows the same for various $\epsilon$ values for
  fixed  $N=70$. The error bars on these
graphs are very small, and hence appear as horizontal lines.}
\end{center}
\end{figure}

In Figure \ref{unscaled.fig} we show results from new simulations,
which plot  $\av{S}$ as a function of $\beta$ for various values of
$N$
 and $\epsilon$. The phase transition is clearly visible even at
$N=30$. For a fixed $\epsilon$, as one varies over $N$
the curves
 show a ``sharpening'' of the transition as $N$ increases.


In~\cite{2dqg} while several values of $N$ and $\epsilon$ were
explored, the most extensive simulations were done only for a few values
of $N$ and $\epsilon$. The existing data was insufficient to deduce scaling behaviour
with any confidence. In order to do so, we have for this work generated a  far more
extensive and comprehensive data set, which we have  analysed along
with the existing data.  We
find that
 plotting $\av{S}$ vs $\beta N$ (see Figure
\ref{rescaledS.fig})
 collapses the transition and the high
temperature curve of
 $\av{S}$, while plotting $\av{S}/N^2$ vs
$\beta N$ collapses the transition
 and the low temperature curve of
$\av{S}$.
 These prescriptions in general give excellent collapse on
either side of the
 phase transition.  Focusing on the high
temperature continuum phase, we find that continuum phase does not
correspond to flat spacetime, but instead has a non-zero,
negative cosmological constant which decreases  linearly with $\beta$,
from the random $\twod $ order at $\beta=0$, up to the phase
transition.
\section{Scaling}
\label{three.sec}

Before embarking on analysing the data, we first lay the ground for
our scaling analysis.
We  are interested in the scaling behaviour of the observable $\av{S}$
from which we can glean the scaling of the specific heat $C$.  In
terms of the free energy  $\beta F=- \ln Z$
\begin{equation}
\label{actionspht}
{ {\beta \av{S} = \beta \frac{\partial (\beta F)}{\partial \beta}, \quad
C= \beta^2(\av{S-\av{S}})^2=-\beta^2 \frac{\partial^2 (\beta F)}{\partial
  \beta^2}.}}
\end{equation}

\subsection{Scaling in the small  and large $\beta$ limit: an analytic argument}

At $\beta=0$, $\av{S}$ is dominated by the ensemble of 2d random
orders~\cite{ezs,winkler} where $\av{S}\sim 4$~\cite{gaussbonnet,bdjs}. Hence there is no scaling with
$N$ at $\beta=0$. Assuming continuity of the partition function for
small enough $\beta$ the density of states will still dominate over
the action. Indeed, Figure \ref{unscaled.fig} indicates that $\av{S}$
is in fact a continuous function of $\beta$, suggesting that the
deviation from flat spacetime and the continuum, if any, should occur
continuously.  If the deviation from flatness is approximately uniform
for small enough $\beta$, with spatial variations in the curvature
being picked up only at larger $\beta$, then one could expect
$S -4 \propto N$. This comes from a direct comparison with the
continuum action for a constant curvature spacetime
$ \frac{1}{16 \pi G} R V$, where $R$ is the scalar curvature, and
$V$ the spacetime volume.  In this case we would expect that for
small enough $\beta$ the scaling exponent is either $0$ or
$1$. Indeed, no other scaling exponent has an obvious continuum or geometric interpretation.

\begin{figure}
\begin{minipage}{0.48\textwidth}
\includegraphics[width=\textwidth]{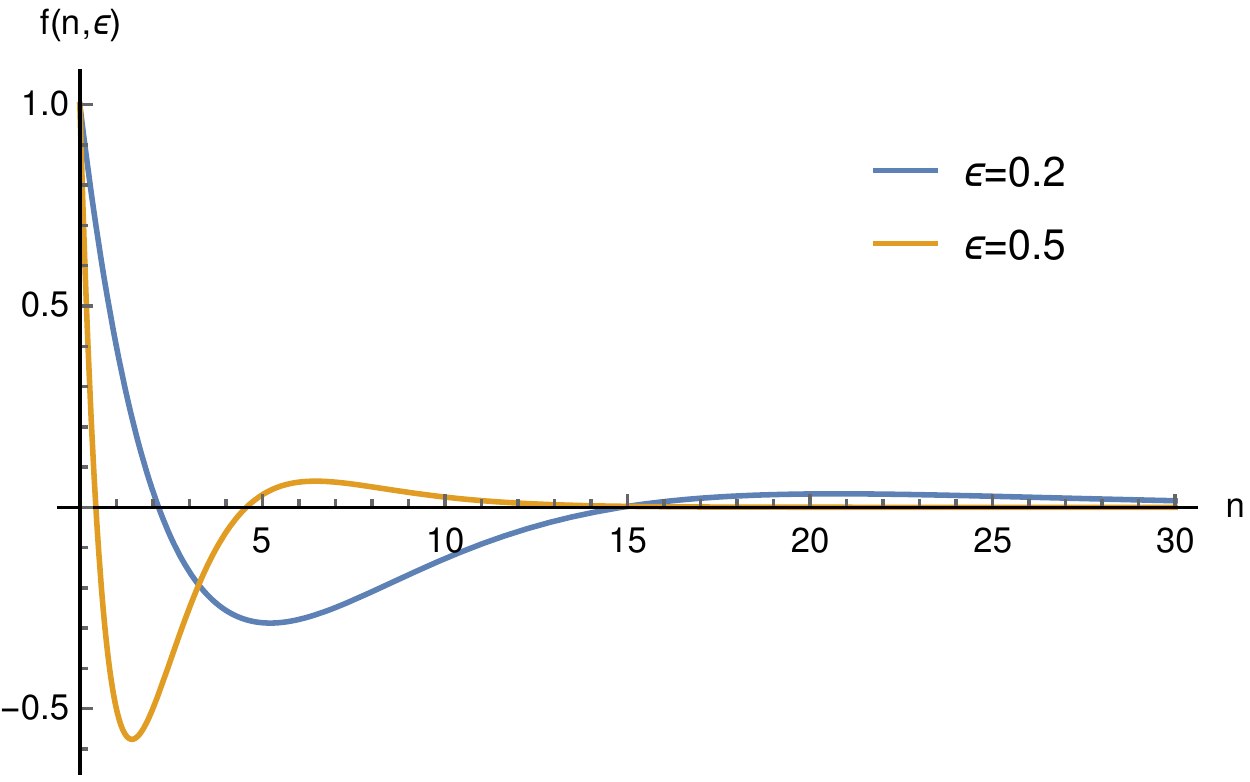}
\caption{The function $f(n,\epsilon)$ smears the contribution to the
  action from different $n$-element intervals. Larger $\epsilon$ means
  less smearing.}
\label{fne.fig}
\end{minipage}\hspace{0.04\textwidth}
\begin{minipage}{0.48\textwidth}
  \centering
  \includegraphics[width=0.7\textwidth]{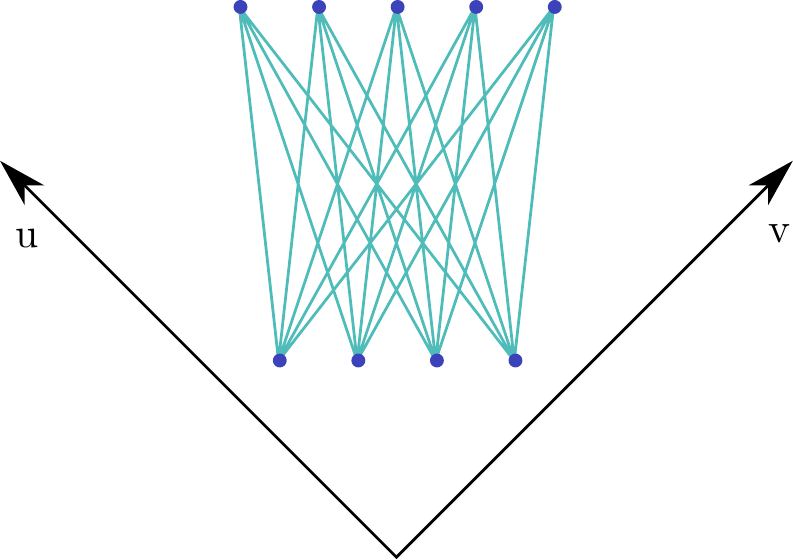}
  \caption{A bilayer causet with each element in the first layer
    linked to each elements in the second layer has the  maximal
    possible number of links.}
  \label{fig.bilayer}
\end{minipage}
\end{figure}

In the opposite limit of large $\beta$ on the other hand, the action
dominates the density of states, and the dominant configuration is the
one with the smallest energy. Now, the BD action (\ref{eq:action2d})
is not positive definite and its sign depends on the details of the
causal set. The function $f(n,\epsilon)$ (Eqn (\ref{fne})) crosses over
from being positive at small $n$, to negative for an intermediate
range of values of $n$ and then goes back to being positive. The
precise location of the cross-over depends on $\epsilon$ as shown in
Figure \ref{fne.fig}.
In particular, $f(n,\epsilon)$ is positive and takes on its largest
value when $n=0$, i.e., for the links. This implies that the lowest
energy causal sets have the the largest number of links. However, this
energetic component has to compete with the density of states at
finite $\beta$.  At small $\beta$, the density of caual sets dominates
the action. Indeed, the $\twod$ random orders do not have the largest
number of links $N_0$ for a given $N$, and neither does the ensemble
at small but non-zero $\beta$~\cite{2dqg}.

The class of $\twod$ orders with the largest number of links for a given
$N$ are bilayer posets, i.e., those with $\frac{N}{2}$ elements in each
of the two layers, and such that every element in a layer is linked to
all the elements in the other layer, as illustrated in Figure \ref{fig.bilayer}. For such posets,
$N_0=\frac{N^2}{4}$, with $N_n=0$ for all $n>0$.
The action has the simple form
$S=4 \epsilon N- 2 \epsilon^2 N^2 $ which scales like
$ \sim N^2$ in the large $N$ limit.


We can now compare the contribution to the partition function from the
random orders $Z_{r}(\beta, N, \epsilon)$ to that from these maximally
connected bilayer posets $Z_b(\beta, N, \epsilon)$. If $\rho_{r,b}(\beta,
N, \epsilon)$ denote the respective density of states
\begin{eqnarray}
\label{dom}
Z_{r}(\beta, N, \epsilon) & \equiv &  \rho_{r}(\beta, N, \epsilon)
                                     \sim N!
\\
Z_b (\beta, N, \epsilon) &\equiv  &  \rho_{b}(\beta, N, \epsilon)
{\rm e}^{+2\beta\epsilon^2 N^2 }
\end{eqnarray}
where we have used the results of~\cite{ezs,winkler} for
$\rho_r (\beta=0, N)$ and normalised by the common factor
${\rm e}^{-4\beta }$. For $\beta \epsilon^2$ large enough, $Z_b$ will
clearly dominate $ Z_r$
for large $N$ irrespective of $\rho_b$.


This analysis  suggests  strongly that the dominant contribution to
the partition function in the large $\beta$ limit comes from bilayer
posets and therefore that  $\av{S} \sim  N^2$ for large $\beta$.

\subsection{Consistency}

The scaling of $\av{S}$ and $C$ must be consistent with the fact that
they are both derived from the free energy $F$ (Eqn
({\ref{actionspht})). In particular, away from the phase transition,
  $\beta \av{S}$ and $C$ must have the same scaling exponents as $\beta F$.

  To begin with, given that $\beta_c$ seems to change with $N$ in
  Figure \ref{unscaled.fig}, let us assume an $N$-dependence of
  $\beta$ and define our first scaling exponent
\begin{equation}
\bb = \beta N^\lambda,
\end{equation}
where $\bb$ is scale independent to leading order in $N$.  Then to
leading order the scaling exponent  for the free energy is given by
\begin{equation}
\label{scalingF}
\beta F\sim  \bb {\bar F}  N^\nu,
\end{equation}
where  $\bbF$  is independent of $N$.  $\nu$ is therefore also the
scaling exponent for $\beta \av{S}$ and $C$.

The scaling of the critical temperature $\beta_c(N,\epsilon)$ with $N$
should give $\lambda$, and the scaling of $\beta \av{S}$,  $\nu$.  In
general, we should expect a different scaling for $\beta
\av{S}$ (or $\beta F$) on either side of the phase transition. Let
$\Pi_\pm$ refer to the phases $\beta<(>) \beta_c$, and $\nu_\pm$ the
respective critical exponents.

While the scaling exponents of $C$ are also $\nu_\pm$ away from the
phase transition, {\it at} the phase transition the scaling depends
on whether the transition is continuous or discontinuous.  For a first
order phase transition the first derivative of the free energy and
hence $\av{S}$ is discontinuous.  Figures \ref{unscaled.fig} suggests
this is a fair guess for our system since as $N$ increases the transition becomes
sharper, and $C$ develops a singularity at the transition, which is
consistent with Figure \ref{sptht.fig}.  Moreover we detect phase coexistence in a small region around the phase transition, Figure \ref
{phasecoexistence.fig}.
  \begin{figure}
    \centering
    \includegraphics[width=0.7\textwidth]{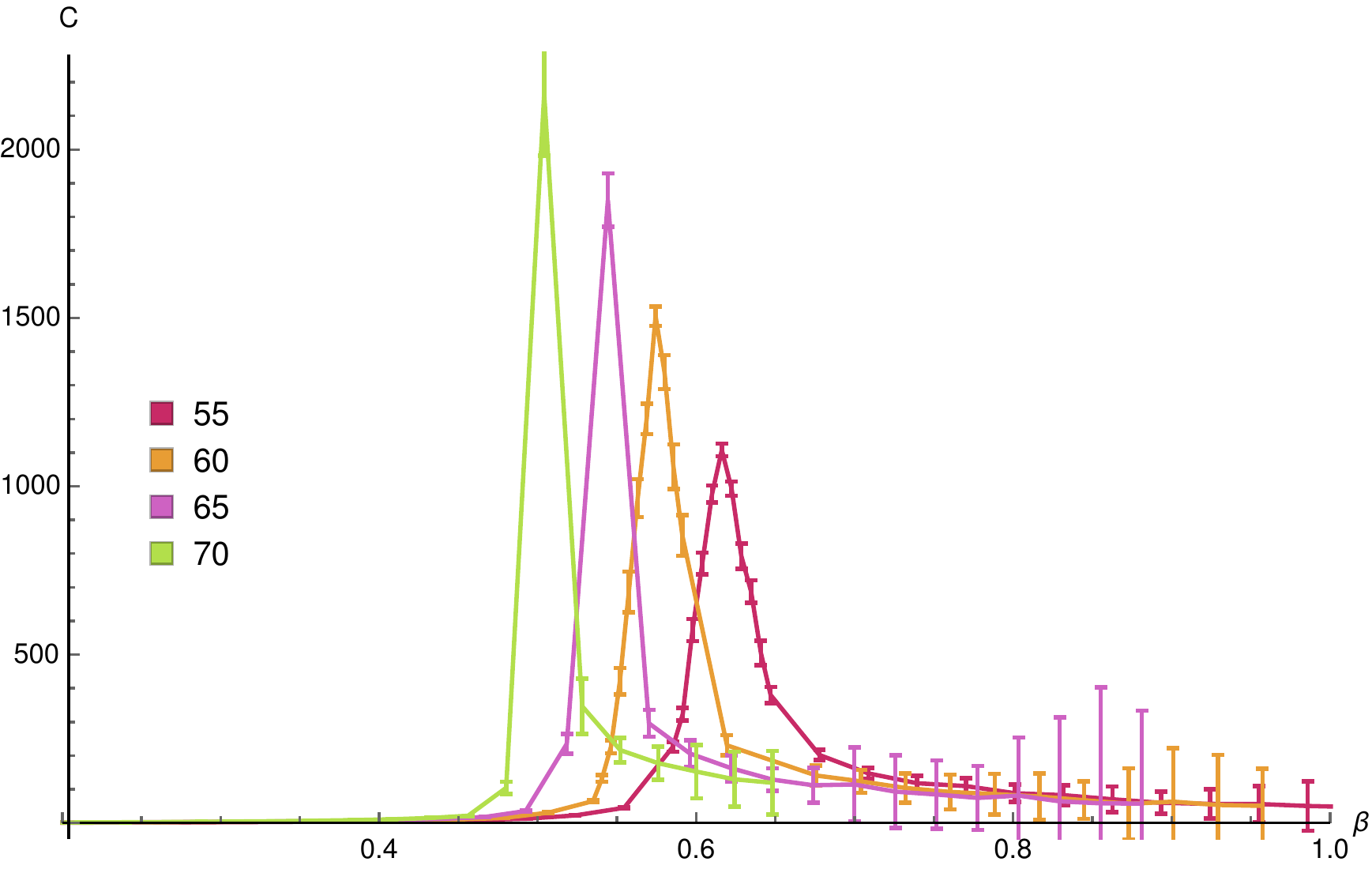}
    \caption{As $N$ increases the peak in the specific heat $C$ grows
      larger and sharper. The plot is for  $\epsilon=0.21$.}
    \label{sptht.fig}
  \end{figure}
\begin{figure}
\centering{
\includegraphics[width=0.6\textwidth]{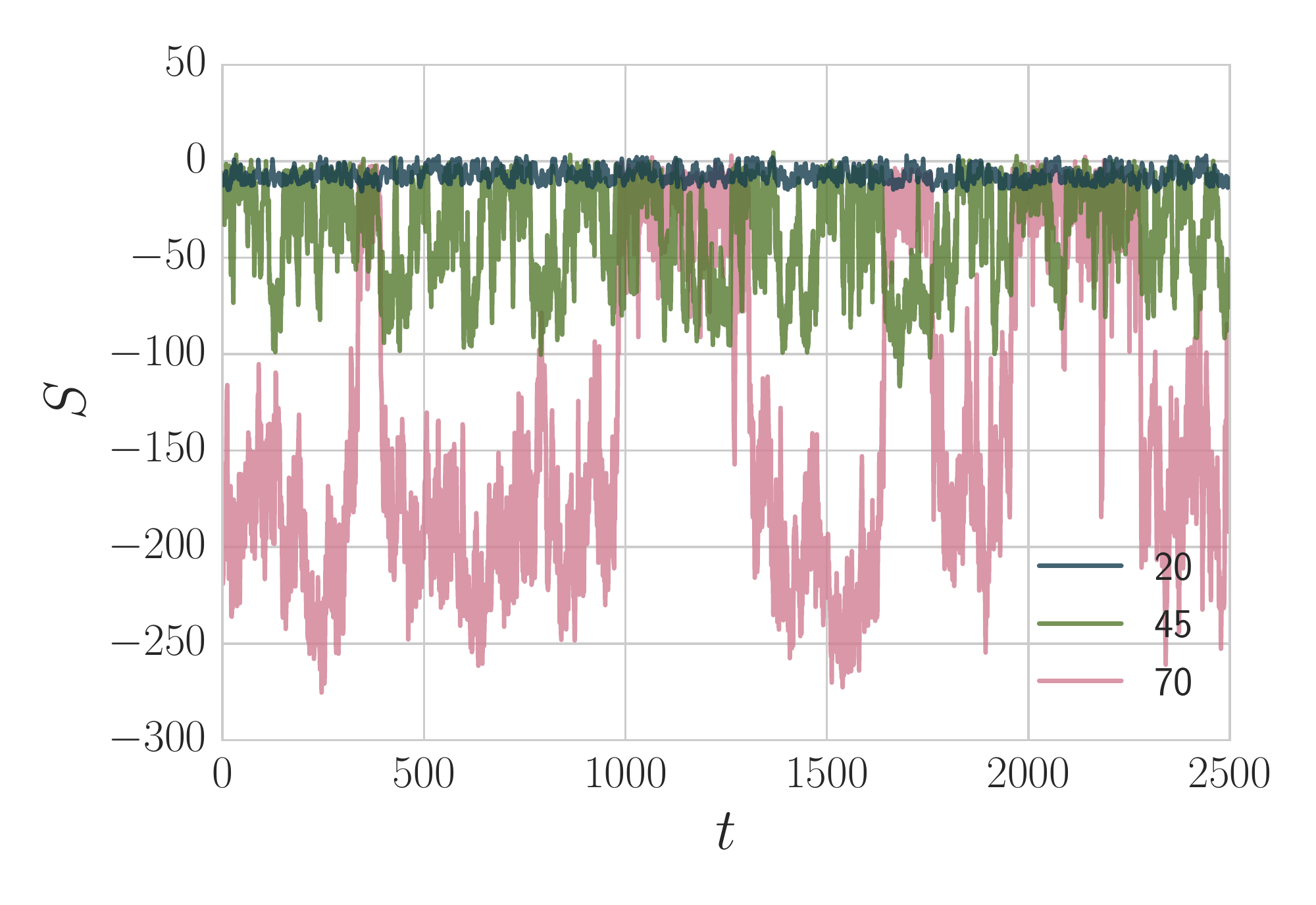}
\caption{The coexistence of  different phases at the phase
  transition for fixed $\epsilon=0.2$,  $N=20,45,70$ with the Markov chain
  jumping  between the two phases. With larger $N$ the difference in
  action of the regions becomes larger and jumps become less frequent,
but more pronounced.}
\label{phasecoexistence.fig}
}
\end{figure}

However, \textit{establishing} a discontinuity or a singularity  in a
finite system requires more than such indications. Indeed,
as pointed out in~\cite{challa},  a finite peak is seen for both first and second order
transitions in finite systems.  Additionally,
phase coexistence occurs  in  finite systems for both types of
transitions,  and result  in double Gaussians in the frequency histograms of the order parameters.  The distinguishing
feature is that the double Gaussians  persist and
become more pronounced in a
first order phase transition  as $N$ is increased, while for a second
order phase transition the two begin to merge. Indeed, this is what
our data shows as we will demonstrate in Section \ref{four.sec}.


The nature of the phase transition affects the scaling of
$C$ with  $N$ at the phase transition. For a second order phase
transition, since the distribution goes over to a single
Gaussian,  the scaling of $C$ is the same as that of $\beta \av{S}$ in
this regime. This also means that the scaling of $\beta \av{S}$
must be the same across the transition.  However, this is not
true for a first order phase transition, where the coexistence of phases is
characterised by two well separated Gaussians, each with a given mean
and variance. The probability distribution for any observable  $x$ at a given
$\beta$ is therefore
\begin{equation}
\label{nong}
  P(x)=p_{+}(\beta)  P_{G}(x,\mu_{+},\sigma_{+},\beta)+p_{-}(\beta) P_G(x,\mu_{-},\sigma_{-},\beta),
  \end{equation}
where  $P_G(x,\mu_\pm,\sigma_\pm,\beta)$ are themselves Gaussians with mean $\mu_\pm$ and standard deviation
$\sigma_\pm$.  $p_\pm(\beta)$ are the relative frequency 
configurations  in $\pi_\pm$ in the ensemble, at a given $\beta$.
 For $\beta<\beta_c$ away from the
transition, $p_-(\beta)$ goes to zero, and simiarly for $\beta
>\beta_c$, again away from the transition, $p_+(\beta)$ goes to
zero so that away  from the phase transition, Gaussianity is
restored.

At (and near) a first order phase transition the deviation from
Gaussianity, Eqn (\ref{nong})  affects the scaling behaviour. For the
action $S$,
\begin{equation}
 \av{S}= p_{+} \av{S_{+}} +p_{-}\av{S_{-}},
\end{equation}
while the specific heat
\begin{equation}
\label{spht-dg}
  C=p_{+}p_{-}(\beta \av{S_{+}} -\beta \av{S_{-}})^2+p_{+}\beta
  C_{+}+p_{-}\beta C_{-}.
\end{equation}
where $\av{S_\pm}$ and $C_\pm$ are the average $S$ and $C$ in
$\Pi_\pm$, respectively.
Given the scaling on either side of the phase transition
\begin{equation}
\beta \av{S_{\pm}}\sim \beta
\av{\bbS_{\pm}} N^{\nu_\pm}
\end{equation}
we expect
\begin{equation}
C_\pm(\beta, N, \epsilon) \sim \bC_\pm (\bb,\epsilon)N^{\nu_\pm}.
\end{equation}
On the other hand, Eqn (\ref{spht-dg}) implies that
\begin{equation}
C(\beta, N, \epsilon) \sim  \bC(\bb, \epsilon) N^{2\nu}.
\end{equation}
where $\nu$ denotes the larger of $\nu_{\pm}$.

\section{Results}
\label{four.sec}

 Our MCMC code was used to generate $\av{S}$ for
the values of $N$ and $\epsilon$
 given in Table {\ref{simdata.tab}}.
For each $N,\epsilon$ there were
 at a minimum, {$35$} values of
$\beta$ that were explored, so that the total number of data
 points
generated was over {$6000$}. Moreover, for each data point
 there
were {$20,000$} sweeps, each sweep having $N(N-1)/2$
 attempted
MCMC moves. These simulations were done on the HPC at the
 Raman
Research Institute.

The first set of simulations were
done over $\beta$ values evenly spread out from $\beta=0$ to the
maximum possible $\beta> \beta_c$, this being determined by
thermalisation times. The next set of simulations was focused on the
critical region. For some
$N,\epsilon$ values we even did a third refinement.   Each set of simulations took several weeks on the RRI
HPC cluster. The aim was to span the parameter space rather than
focus on fixed  $N$ and $\epsilon$.
\begin{table}
  \caption{Values of $N$ and $\epsilon$ used in our analysis}
  \label{simdata.tab}
  \centering
  \begin{tabular}{l | l l l l l l l l l l }
    \toprule
    N & $\epsilon$ & & & & & & & & &  \\ \midrule
   30 & 0.1 & 0.11 & 0.12 & 0.13 & 0.14 & 0.15 & 0.16 & 0.17 & 0.18 & 0.19 \\
      & 0.2 & 0.21 & 0.22 & 0.23 & 0.24 & 0.25 & 0.3  & 0.35 & 0.4  & 0.5 \\
   35 & 0.1 & 0.11 & 0.12 & 0.13 & 0.14 & 0.15 & 0.16 & 0.17 & 0.18 & 0.19 \\
      &0.2 & 0.21 & 0.22 & 0.23 & 0.24 & 0.25 & 0.3 & 0.35 & 0.4 & 0.5 \\
   40 & 0.1 & 0.11 & 0.12 & 0.13 & 0.14 & 0.15 & 0.16 & 0.17 & 0.18 & 0.19 \\
   & 0.2 & 0.21 & 0.22 & 0.23 & 0.24 & 0.25 & 0.3 & 0.35 & 0.4 & 0.5 \\
   45 & 0.1 & 0.11 & 0.12 & 0.13 & 0.14 & 0.15 & 0.16 & 0.17 & 0.18 & 0.19 \\
   & 0.2 & 0.21 & 0.22 & 0.23 & 0.24 & 0.25 & 0.3 & 0.35 & 0.4 & 0.5 \\
   50 & 0.1 & 0.11 & 0.12 & 0.13 & 0.14 & 0.15 & 0.16 & 0.17 & 0.18 & 0.19 \\
   & 0.2 & 0.21 & 0.22 & 0.23 & 0.24 & 0.25 & 0.3 & 0.35 & 0.4 & 0.5 \\
   55 & 0.1 & 0.11 & 0.12 & 0.13 & 0.14 & 0.15 & 0.16 & 0.17 & 0.18 & 0.19 \\
   & 0.2 & 0.21 & 0.22 & 0.23 & 0.24 & 0.25 & 0.3 & 0.35 & 0.4 & 0.5 \\
   60 & 0.1 & 0.11 & 0.12 & 0.13 & 0.14 & 0.15 & 0.16 & 0.17 & 0.18 & 0.19 \\
   & 0.2 & 0.21 & 0.22 & 0.23 & 0.24 & 0.25 & 0.3 & 0.35 & 0.4 & 0.5 \\
   65 & 0.1 & 0.11 & 0.12 & 0.13 & 0.14 & 0.15 & 0.16 & 0.17 & 0.18 & 0.19 \\
   & 0.2 & 0.21 & 0.22 & 0.23 & 0.24 & 0.25 & 0.3 & 0.35 & 0.4 & 0.5 \\
   70 & 0.1 & 0.11 & 0.12 & 0.13 & 0.14 & 0.15 & 0.16 & 0.17 & 0.18 & 0.19 \\
   & 0.2 & 0.21 & 0.22 & 0.23 & 0.24 & 0.25 & 0.3 & 0.35 & 0.4 & 0.5 \\
   75 & & 0.11 &  \text{} & \text{} & \text{} & \text{} & \text{} & \text{} & \text{} & \text{}\\
   80 & & 0.11 & 0.12 & 0.13 & 0.14 & 0.15 & 0.16 & 0.17 & 0.18 & 0.19  \\
   90 & & 0.11 & 0.13 & 0.14 & 0.15 & \text{} & \text{} &\text{} & \text{}\\ \bottomrule
 \end{tabular}
\end{table}

Importantly, our data shows that the
{\it qualitative}  behaviour is unchanged as $N$ and $\epsilon$  are varied
with the appearance of a phase transition from  one
distinct phase into another  as $\beta$ is varied.  Figure \ref{unscaled.fig}
uses this new data giving $\av{S}$ as a function of $\beta$ for both a
fixed $\epsilon$($=0.21$)  and varying $N$,  and also for fixed $N$($=70$) and
varying $\epsilon$. From these it is apparent that for fixed $\epsilon$
 the critical inverse temperature  $\beta_c$ decreases  with $N$, and similarly for fixed $N$,
 it decreases with $\epsilon$.  These graphs show a strong hint
 of scaling, but it is clear that this is non-trivial.  A seeming
 worry is that as   $N$ increases,  $\beta_c $ goes to
 $0$, so  that the phase transition appears to vanish in the
 asymptotic regime. However, we now show that when  rescaled as  $\bb=\beta
 N$, the rescaled   $\bb_c$  rapidly converges to a fixed value with $N$.

We note that  in order for the BD action to yield the right
continuum approximation, $\epsilon $ must be large enough for a given
$N$, otherwise the non-locality scale $l_p/\sqrt{\epsilon}$ will
exceed the IR cut off. In this case, the BD action will not yield the
Einstein Hilbert action in the continuum approximation and hence is
not quantum gravity as we have defined it.
Thus there exists a smallest $\epsilon_0$ for
every $N$. The expectation is that $\epsilon_0(N) \sim N^{-c} $ for $
c>0$, so that $\epsilon_0(N) \rightarrow 0$ as $N\rightarrow \infty$.
In analysing our data this cut-off was taken into account.
\subsection{Scaling of $\beta$ }

We estimate $\beta_c$ as the location of the maxima of the
(simulated) value of the specific heat $C$. Clearly, the real maximum
 could lie at a value of $\beta$ which we did not simulate and hence
 it is important to assess the error $\Delta \beta_c$. We take this to be  the largest interval on the  $\beta$ axis which
contains  $\beta_c$, such that the
errors in $C$ of  the end point values of the interval  overlap with
the errors in $C$ at $\beta_c$. We illustrate this in
\ref{betacerror.fig}.
\begin{figure}[b!]
\centering{
 \includegraphics[width=0.7\textwidth]{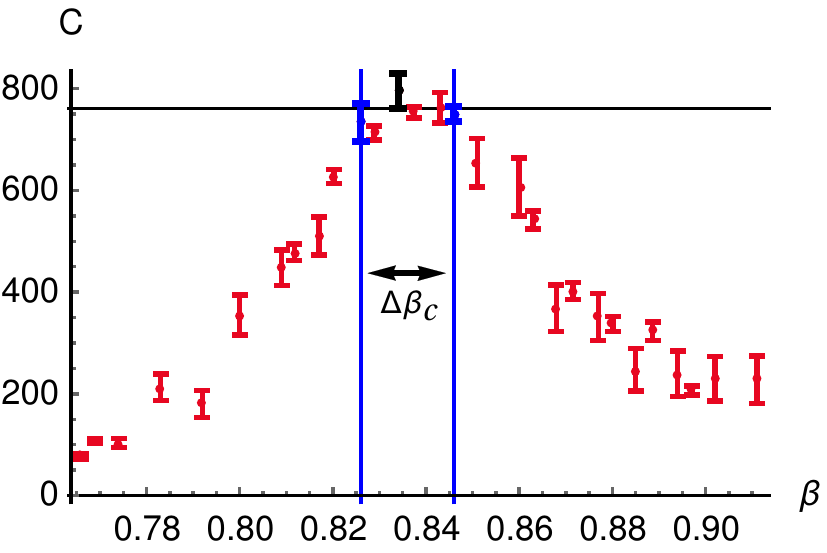}
\caption{\label{betacerror.fig} We  take the pseudocritical point $\beta_c$ to
  be the location of the maxima of $C$. To estimate the error $\Delta
  \beta_c$ we compare the error  in $C_{\mathrm{max}}$ at $\beta_c$ with that of neighbouring
  points. The  two $\beta$ values on either side of $\beta_c$
  which are furthest away from it  and such that their errors in $C$ overlap
  with that at $C_{\mathrm{max}}$ then determine  $\Delta
  \beta_c$.}}
\end{figure}



\begin{figure}
\centering{
\includegraphics[width=0.75\textwidth]{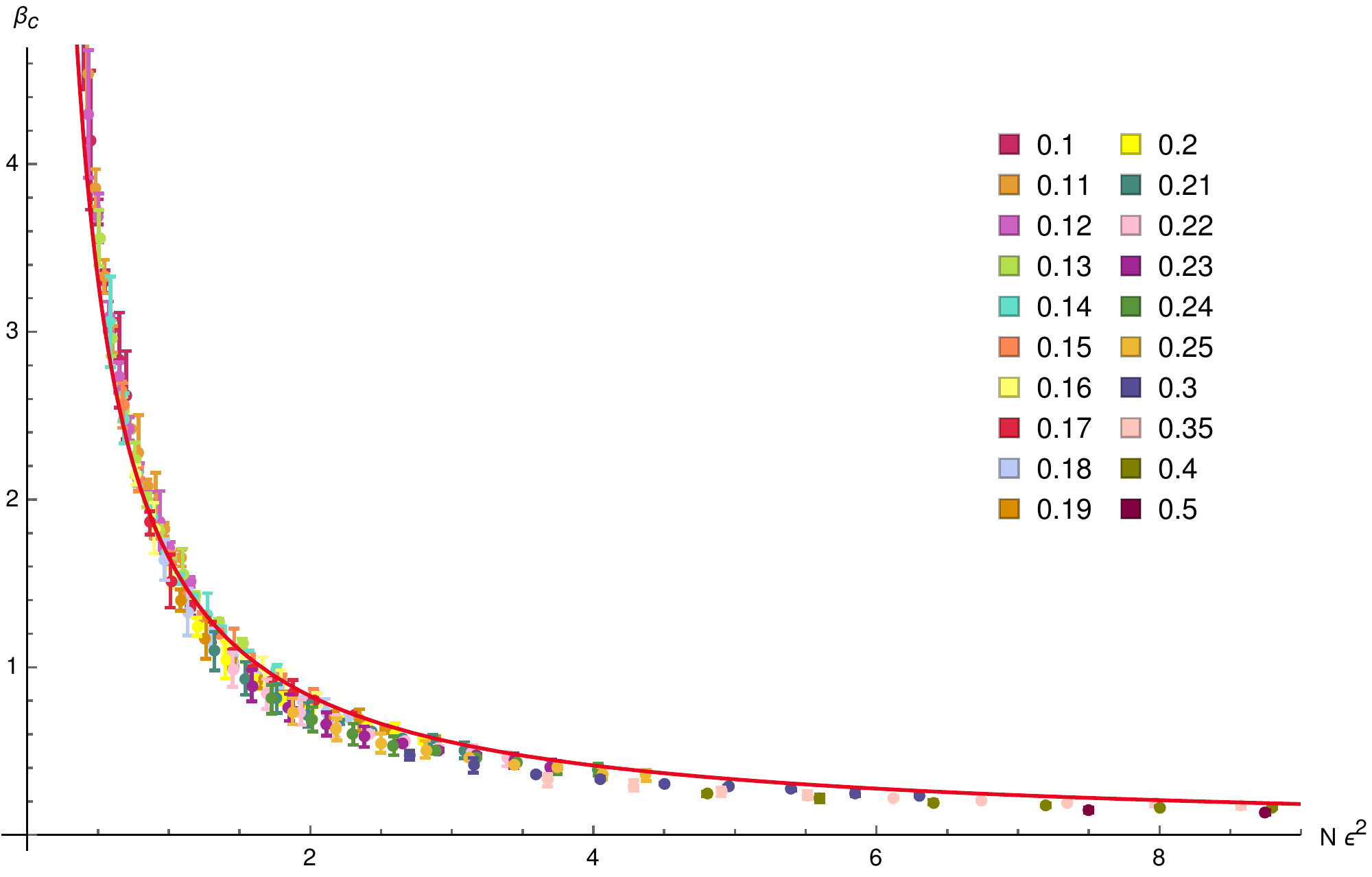}
\caption{
We show $\beta_c$ against $N\epsilon^2$, and see that the data collapses very well for all values of $N,\epsilon$.
\label{betaccoll}}}
\end{figure}
Our data suggests a scaling  $\beta_c \sim \frac{1}{N\epsilon^2}$
as shown in Figure \ref{betaccoll}.  Assuming
\begin{equation}\label{eq:betac}
\beta_c= \frac{b(\epsilon)}{N}+\frac{c(\epsilon)}{N^2}+ O(\frac{1}{N^3}),
\end{equation}
we find fits for $b(\epsilon)$ and $c(\epsilon)$ as shown in Figs
\ref{bcfits.fig},
\begin{equation}
b(\epsilon)= \frac{1.66(\pm 0.03)}{\epsilon^2}, \quad c(\epsilon)=\frac{4.09  (\pm 0.50) }{\epsilon^3} -  \frac{27.77(\pm
2.45)}{\epsilon^2}
\end{equation}
\begin{figure}
\includegraphics[width=0.48\textwidth]{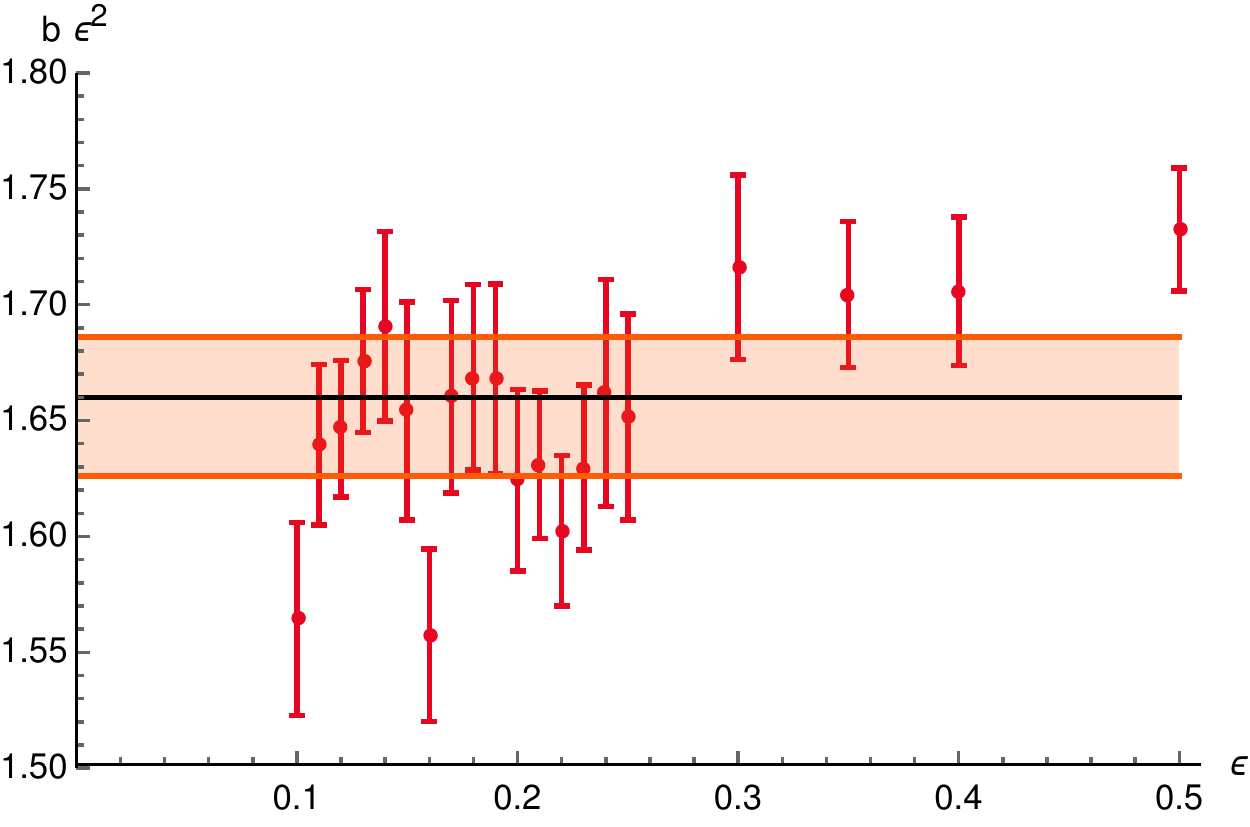}
\hfill
\includegraphics[width=0.48\textwidth]{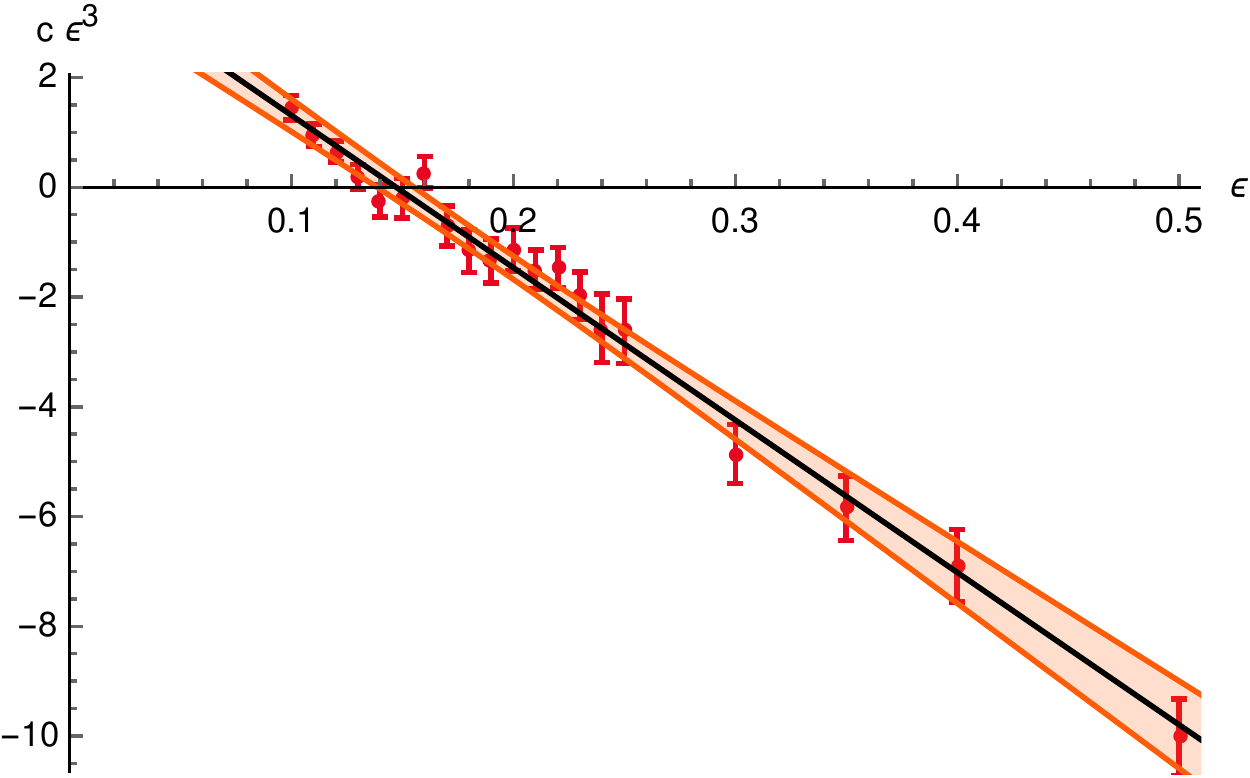}
\caption{\label{bcfits.fig} Best fits for $b,c$ as functions of $\epsilon$.}
\end{figure}
Using these estimates for $b(\epsilon)$ and $c(\epsilon)$, we find
very good fit with the data as shown in Figure  \ref{fitbc.fig}.
\begin{figure}
\centering{
\includegraphics[width=0.75\textwidth]{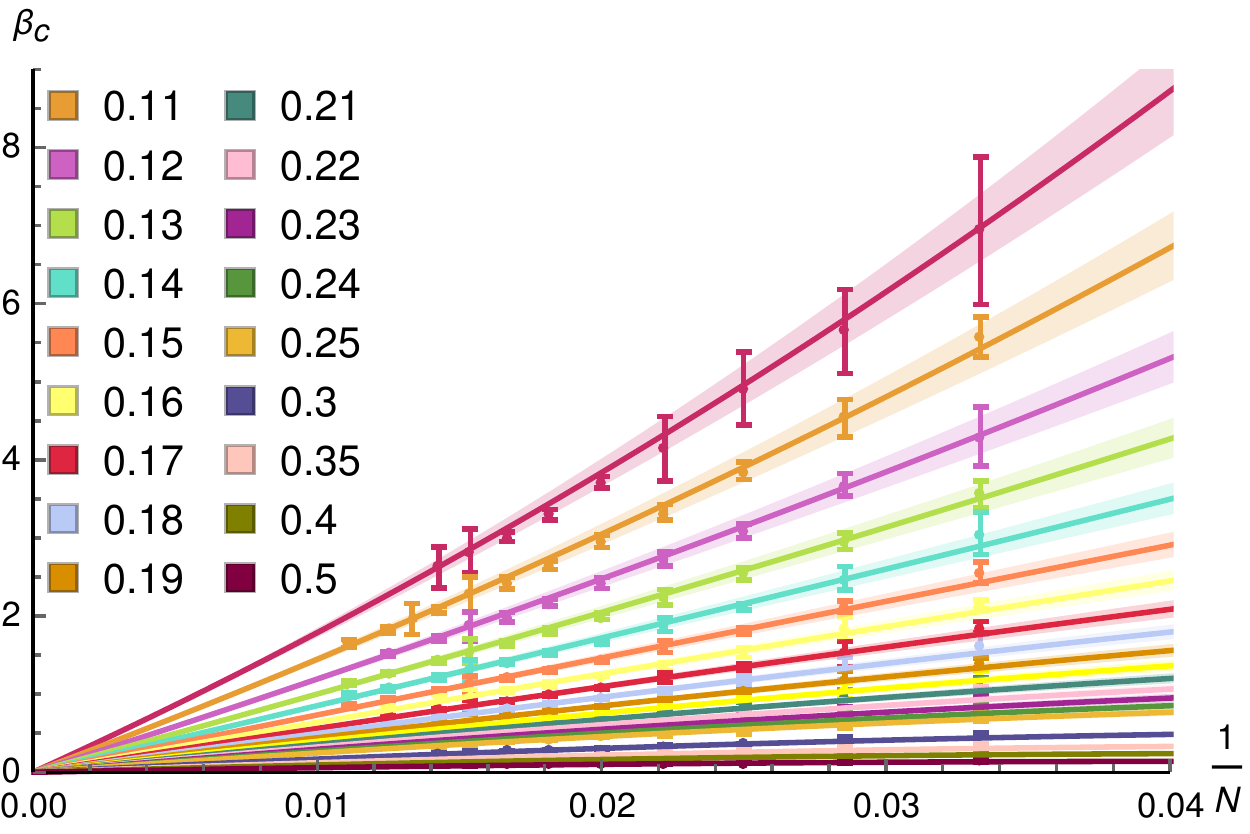} 
\caption{\label{fitbc.fig}Plotting $\beta_c$ vs $\frac{1}{N}$ shows a clear linear dependence. The lines are plotted using equation \eqref{eq:betac}, and the shaded region shows the $99\%$ confidence interval.}
}
\end{figure}
Thus, to leading order, we can define the scale invariant temperature
$\bb \sim \beta N$. Replotting $\av{S}$ v/s $\bb$ we  find a strong  convergence to $\bb_c$ as $N$ increases
as shown in Figure \ref{rescaledS.fig}.
\begin{figure}
\centering{
\includegraphics[width=0.75\textwidth]{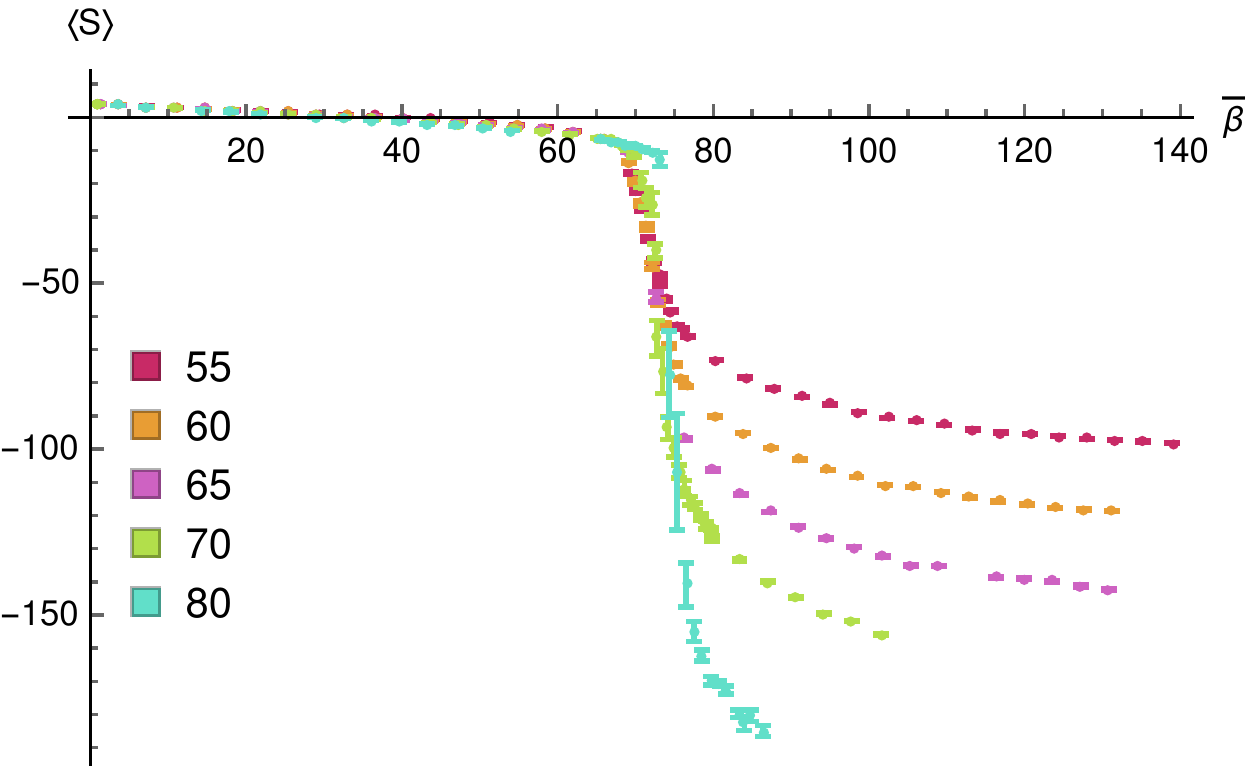}
\caption{\label{rescaledS.fig} This figure shows how for $\epsilon
  =0.15$ the rescaled critical temperature $\bb=\beta N$ converges for
  larger $N$. Again, the error bars are very small, and show up as
  lines in the graph.}
}
\end{figure}

This yields  the first exponent $\lambda=-1$.

\subsection{Scaling of $\av{S}$}

We now examine the scaling of $ \av{S}$ in the
phases $\Pi_\pm$ to estimate the  scaling exponents $\nu_\pm$.

In the continuum phase $\Pi_-$ region we see that
rather than staying  constant as one might expect for the $\twod$ random order,
$\av{S}$, drops away from  $0$ with a linear dependence
on $\bb$, 
as shown in Figure \ref{largeNzoom.fig}.
\begin{figure}
\centering{
 \includegraphics[width=0.75\textwidth]{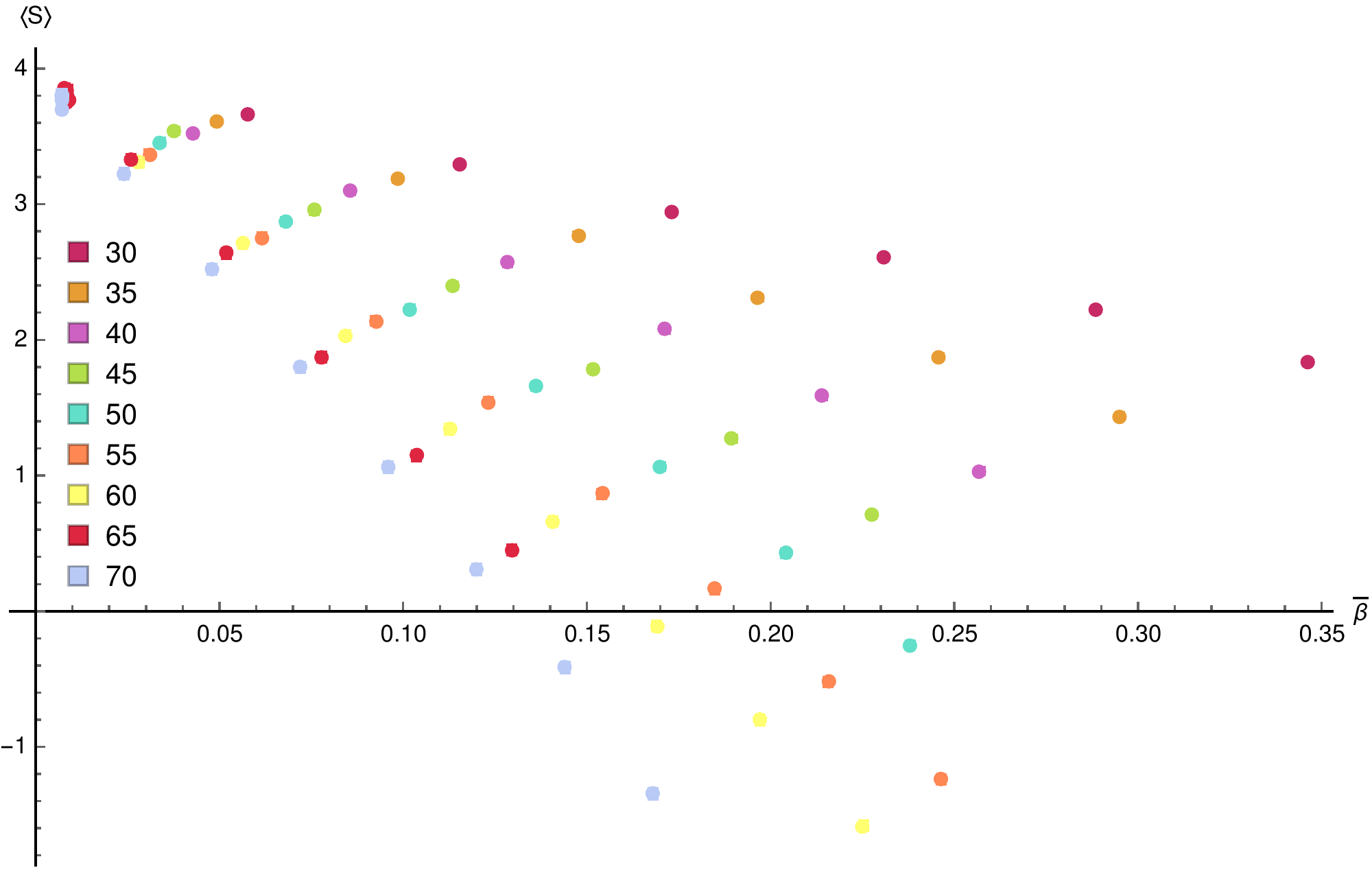}
\caption{\label{largeNzoom.fig} $\av{S}$ as a function of $\bb$
  before the phase transition and the linear best fits for $\epsilon=0.21$.
}
}
\end{figure}
To leading order we guess that
\begin{equation}\label{eq:prePTfit}
\av{S}-4= b^-(N,\epsilon) \bb,
\end{equation}
with
\begin{equation}
b^-(N,\epsilon)=b_1^-(\epsilon)N+b_0^-(\epsilon)
\end{equation}
%
\begin{figure}
\centering{
 \includegraphics[width=0.48\textwidth]{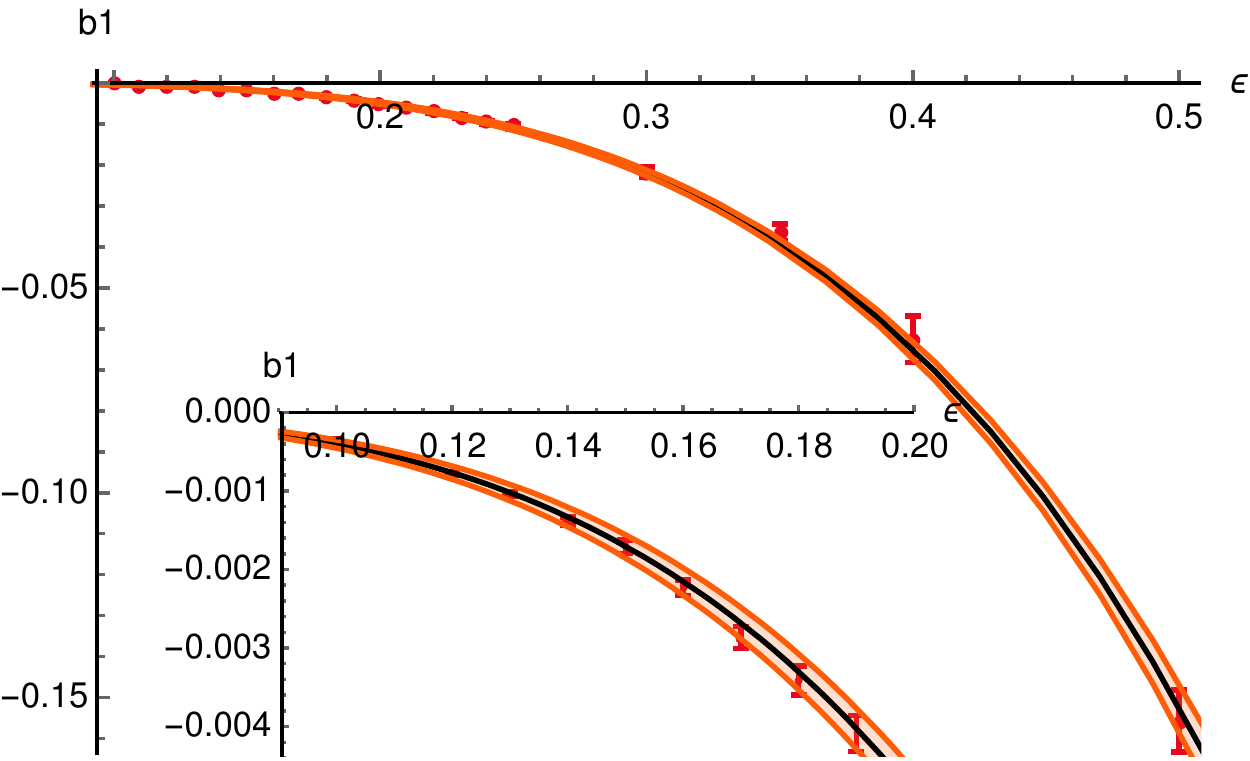} \hfill
\includegraphics[width=0.48\textwidth]{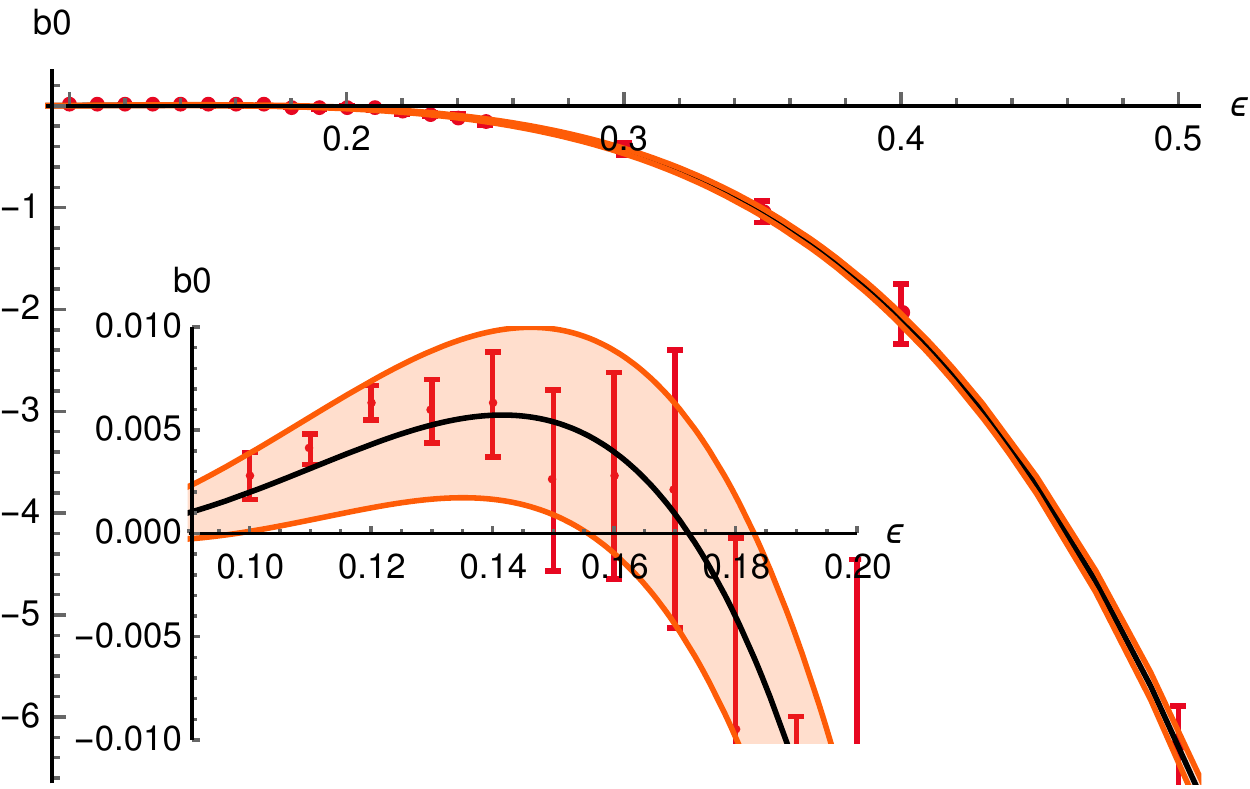}
\caption{\label{Fitsb0b1PT.fig}Best fits  and confidence regions for $b_1(\epsilon),b_0(\epsilon)$. The inlays show a zoom for the small $\epsilon$ region.}
}
\end{figure}
In Figures \ref{Fitsb0b1PT.fig} we find fits for $b_0^-(\epsilon) $
and $ b_1^-(\epsilon)$  and find that
\begin{eqnarray}
  b_1(\epsilon) &=& - 0.20 (\pm 0.06) \epsilon^3 -2.04 (\pm 0.18) \epsilon^4
  \nonumber \\
b_0(\epsilon) &=& 2.09 (\pm 0.55)  \left(\epsilon - 0.07 (\pm 0.01)\right)^2 - 190.50 (\pm
15.46) \left(\epsilon -0.07 (\pm 0.01 )\right)^4 \;.
\end{eqnarray}
The plots also show the $99 \%$ confidence interval for the
fits. In Fig \ref{fig:totalcollapse_beforePT} we divide  the average action
by the best fit function to show the goodness of the fit.

To leading order, then, $\av{S} \sim N$.
This is interesting since for the 2d random orders, the action does not
scale with $N$, i.e., $\av{S} \sim 4$ for all $N$.  Therefore, it is
evident that the ``continuum'' phase is {\it not} dominated by flat
spacetime for $\bb>0$.  As we will discuss in Section \ref{five.sec} this
scaling is consistent with a constant curvature spacetime of
negative curvature, i.e., $\mathrm{adS}_2$. We  show further support for
this, which suggests that $\twod$ CST has a dynamical mechanism for
generating a cosmological constant.

\begin{figure}
\centering{
  \includegraphics[width=0.75\textwidth]{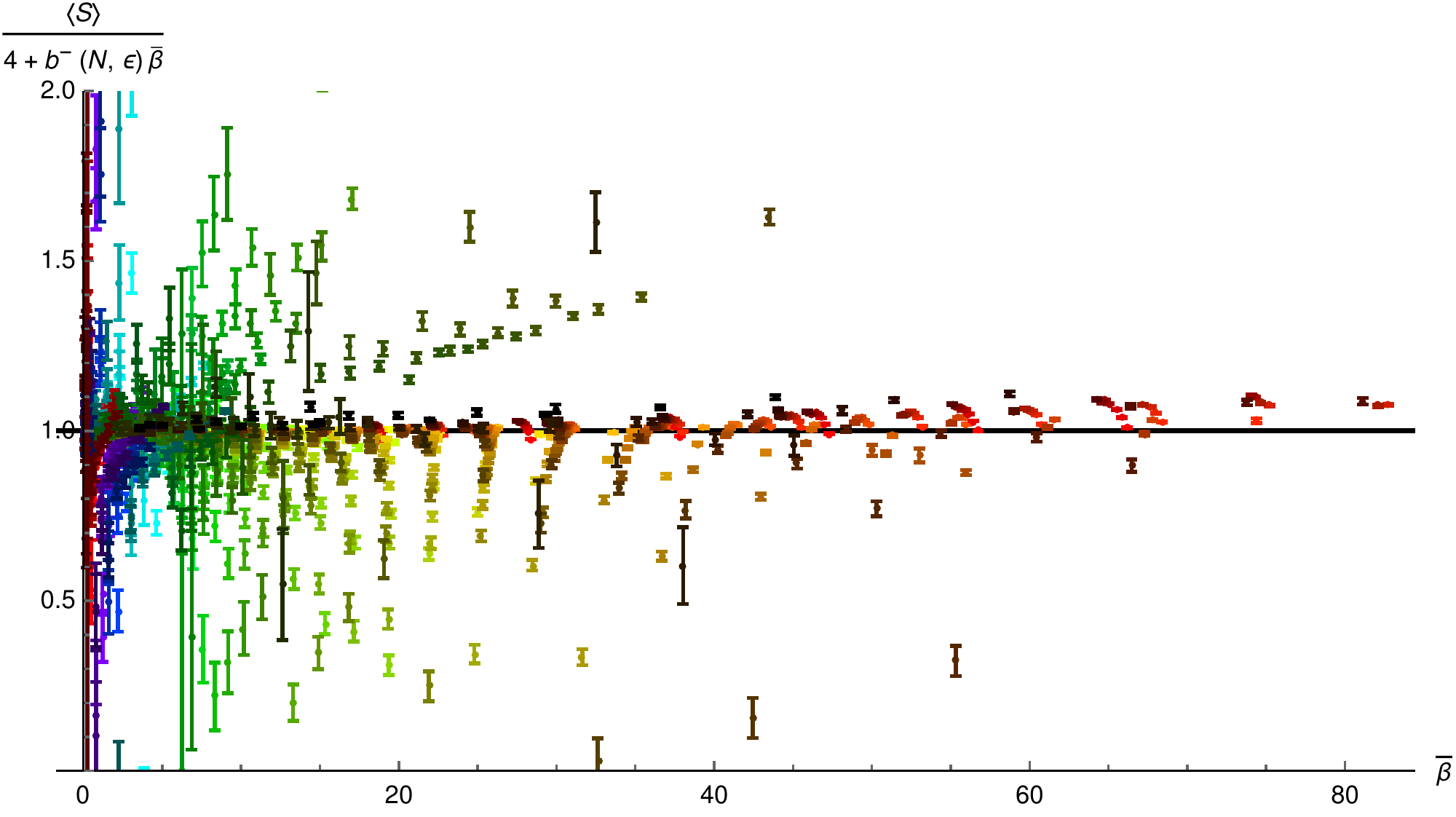}
  \caption{
  Dividing the average action by the best fit function for the region before the phase transition \eqref{eq:prePTfit} shows a very good fit for all ranges of $N,\epsilon$.
  The color shows the value of $\epsilon$ (with $\epsilon$ decreasing from left to right), while the brightness indicates $N$, with darker dots indicating larger $N$ values.}
  \label{fig:totalcollapse_beforePT}}
\end{figure}

Since $\beta$ itself scales as $N^{-1}$, this
means that $\beta \av{S} \sim N^0$, or that $\nu_-=0$ as is clear from
the convergence of the larger values of $N$ shown in Figure \ref{bSbNPT.fig}.
\begin{figure}
\centering{
 \includegraphics[width=0.75\textwidth]{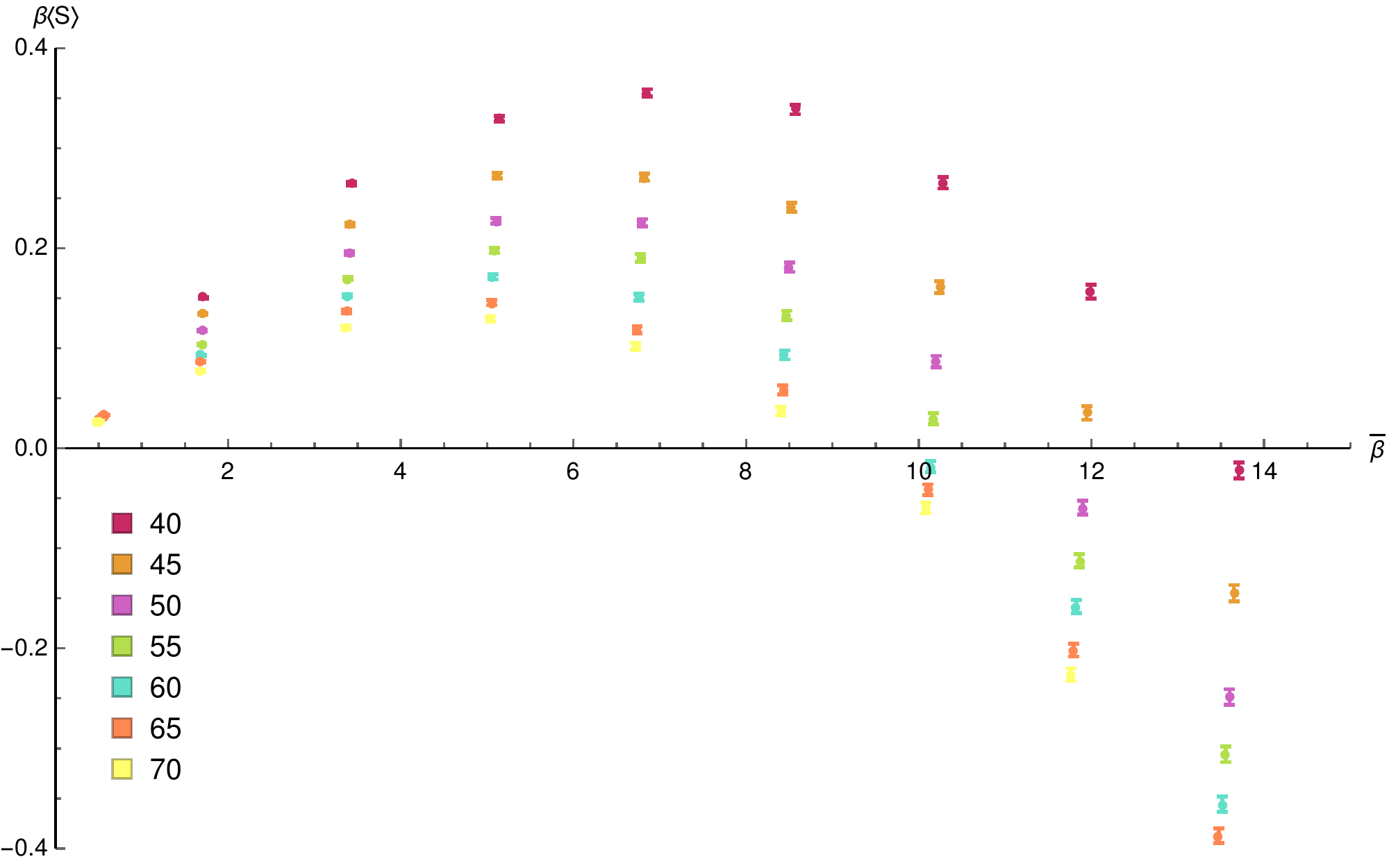}
\caption{\label{bSbNPT.fig} In the phase $\Pi_-$, $\beta \av{S}$ shows a convergence for large $N$.}
}
\end{figure}
We can do more in this case. To leading order since
\begin{equation}
-\bb \, \,\frac{\partial }{\partial
  \bb} {\ln Z}\sim b_1^-(\epsilon) \bb^2
\end{equation}
we see that the free energy $\beta F$ is scale invariant and hence non-extensive.
This expression is in fact so simple, we can integrate
it  to find the partition function
\begin{equation}
\ln \frac{Z}{Z_0} \sim  -\int_0^{\bb} b_1^-(\epsilon) \bb' d\bb' =
-\frac{1}{2} b_1^-(\epsilon)\bb^2.
\end{equation}
Here, $Z_0$ is the partition function at $\beta=0$ which to leading order
in $N$,  is given by the density of states $N!$ for  the $\twod$
random orders  (Eqn \ref{dom}).

At the other end, for $\beta>\beta_c$,
$\av{S}$ changes with $N$, becoming more and more negative
as {$N$} increases  as shown in Figure \ref{largebetaN.fig}(a). In
Figure \ref{largebetaN.fig}(b) we  plot $\beta \av{S}/N$ with $\bb$ and find a
linear behaviour with $\bb$ which converges with increasing $N$.
\begin{figure}
\begin{center}
\includegraphics[width=0.49\textwidth]{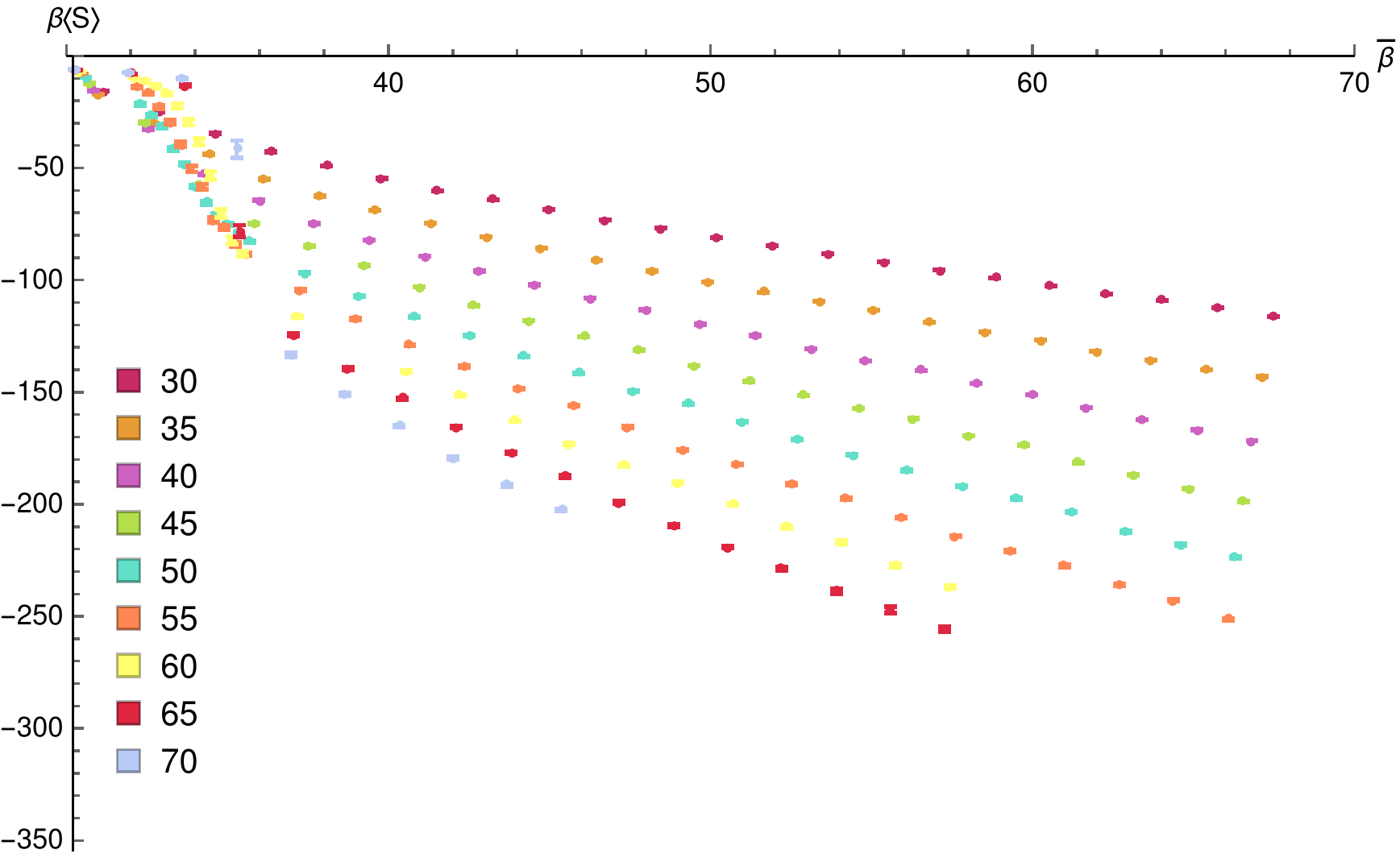}
\includegraphics[width=0.49\textwidth]{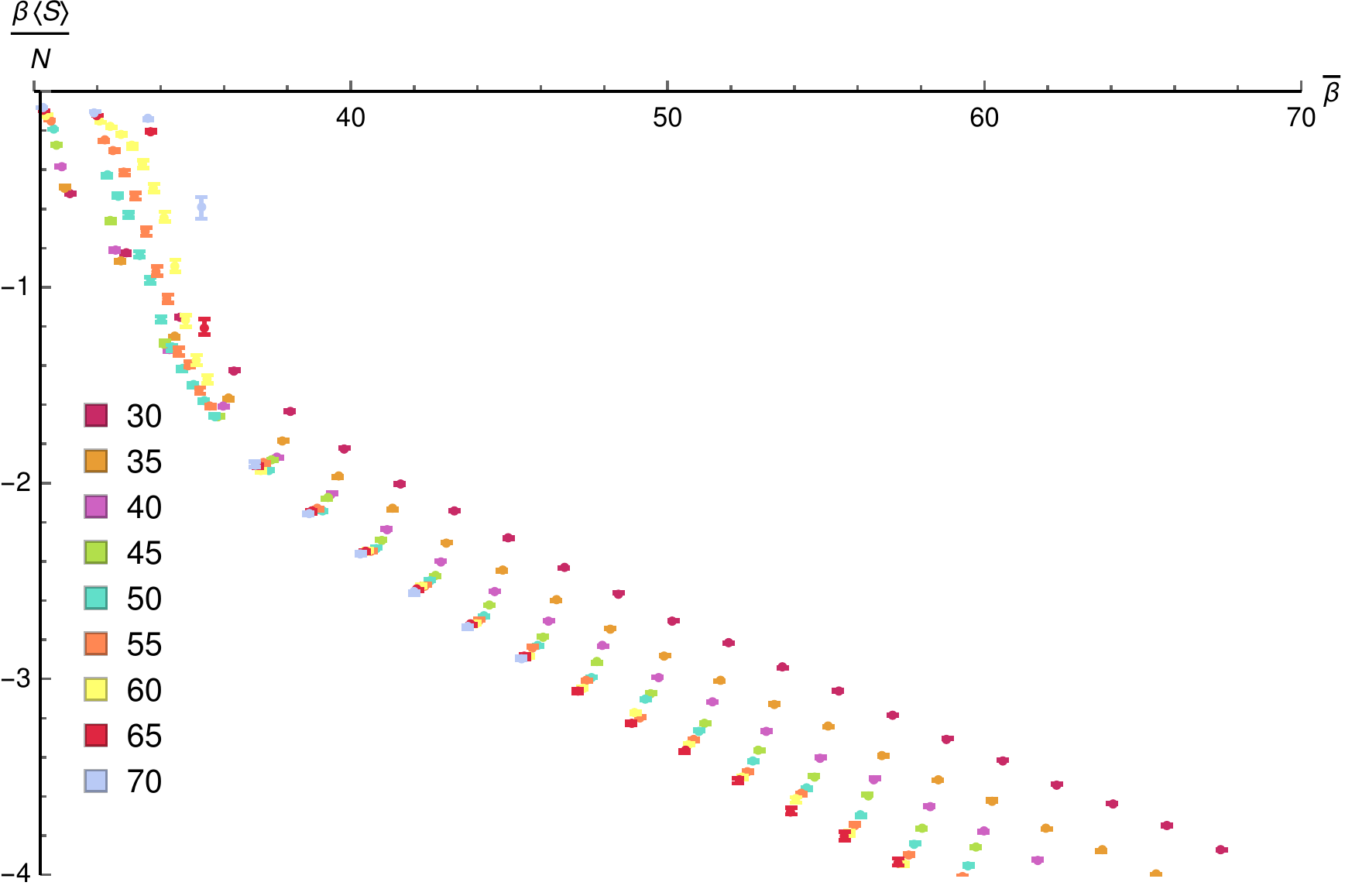}
\end{center}
\caption{\label{largebetaN.fig} In the phase $\Pi_+$, we see $\beta\av{S}$
  behaves linearly with $\bb$  in the figure on the left. On  the right
  we show it rescaled as $\beta \av{S}/N$ which shows convergence for large $N$.}
\end{figure}
Modelling the linear behaviour as
\begin{equation}\label{eq:pastPTfit}
{\beta \av{S}} = a^+(N,\epsilon) + b^+(N,\epsilon) \bb,
\end{equation}
we let
\begin{equation}
a^+(N,\epsilon)=  a_0^+(\epsilon)+a_1^+(\epsilon) N, \quad
b^+(N,\epsilon)=b_0^+(\epsilon) + b_1^+(\epsilon) N.
\end{equation}
In Figures \ref{APTfitsa.fig},\ref{APTfitsb.fig} we show the best fits
for these functions and find that
\begin{figure}
\begin{center}

\includegraphics[width=0.45\textwidth] {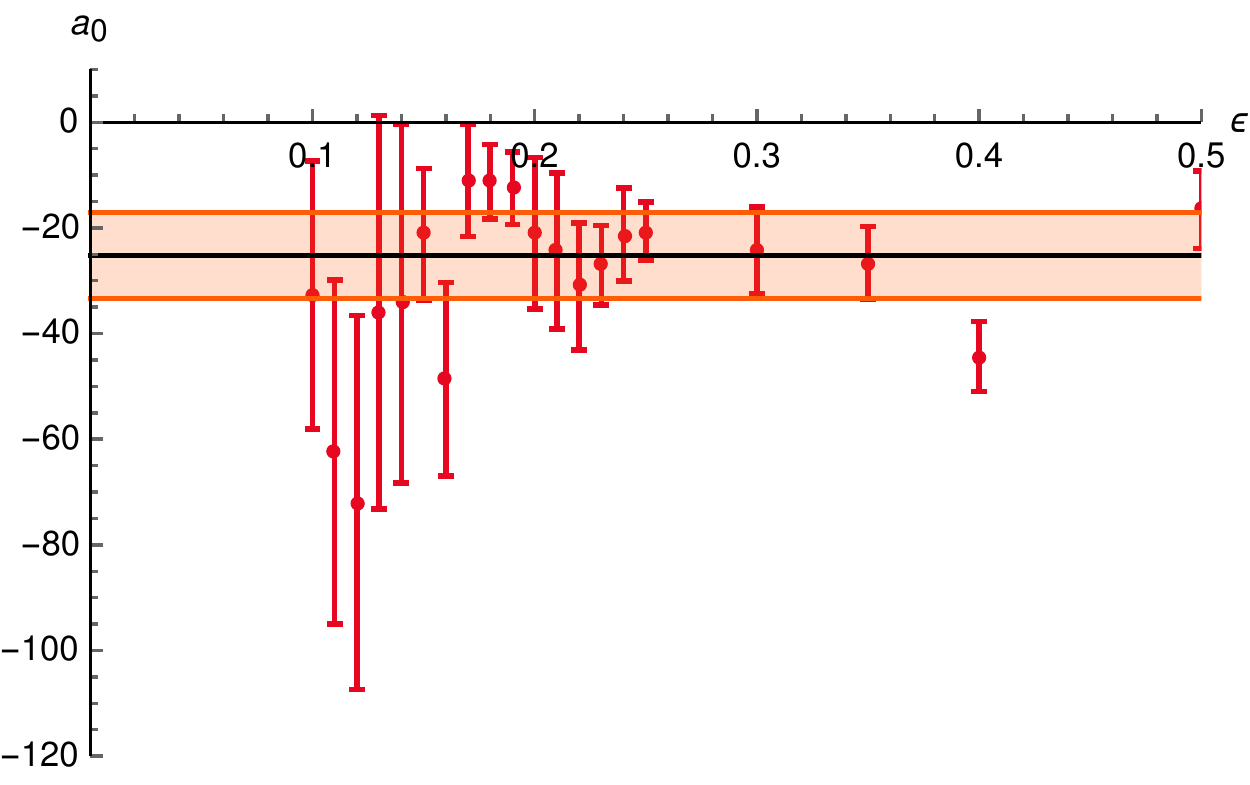} \hfill
 \includegraphics[width=0.45\textwidth] {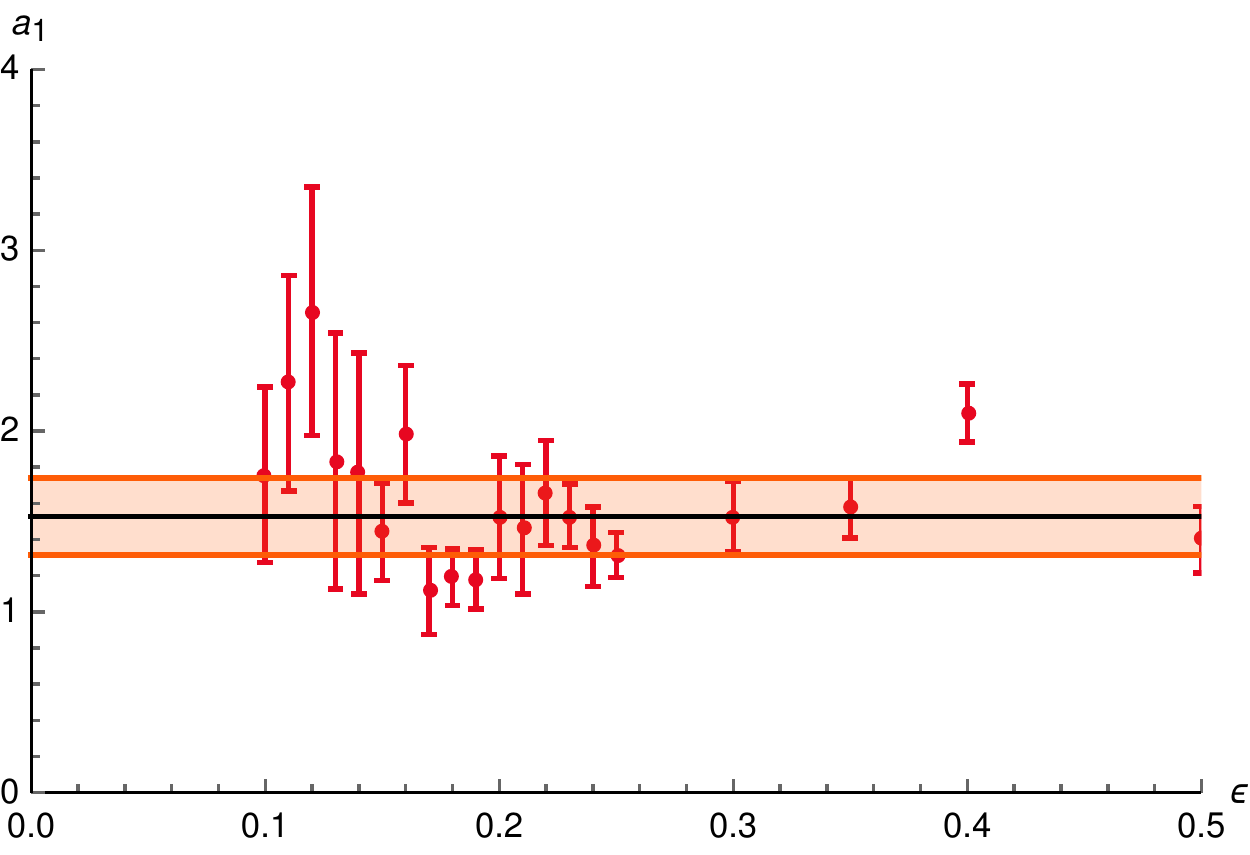}
\end{center}
\caption{\label{APTfitsa.fig}  Best fits for $a_0$ and $a_1$ as
  functions of $\epsilon$ and 99 $\%$ confidence intervals.}
\end{figure}
 \begin{figure}
\begin{center}

\includegraphics[width=0.45\textwidth] {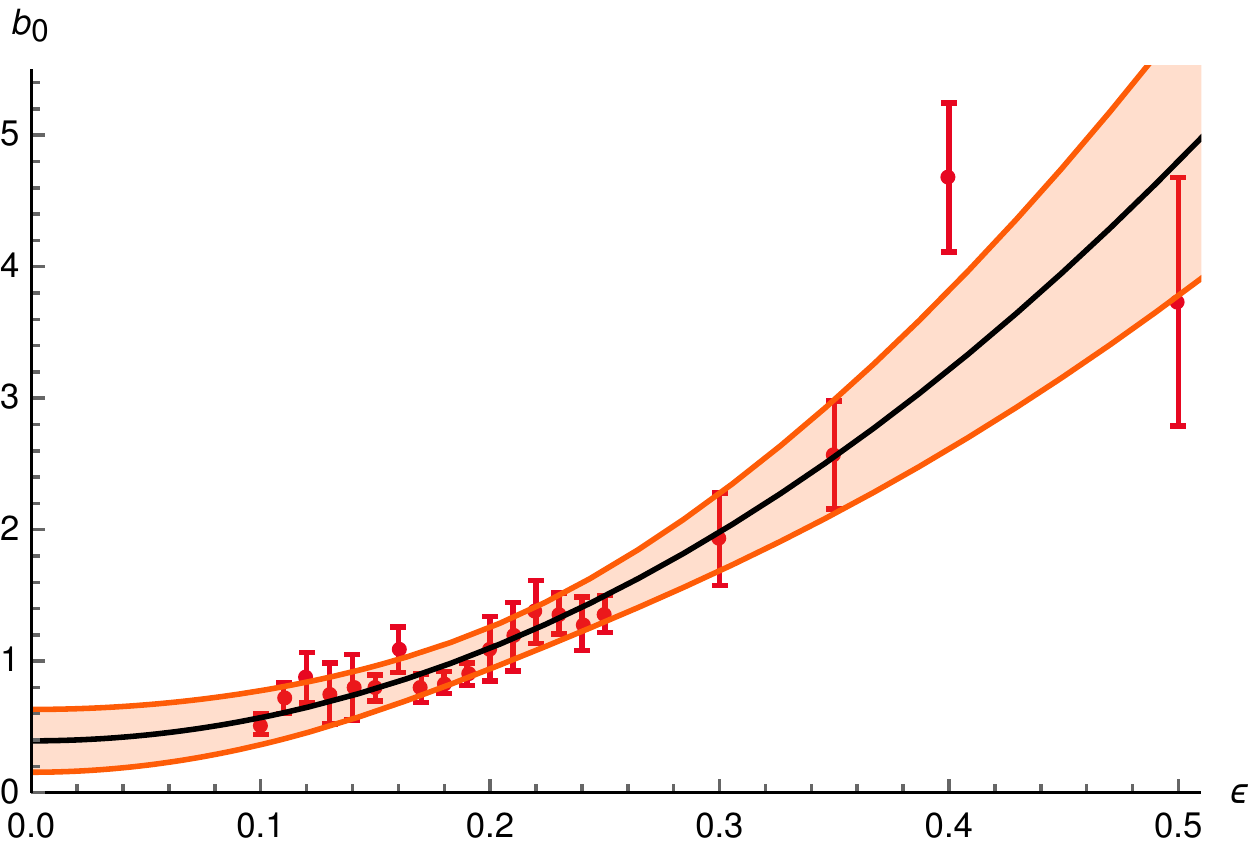} \hfill
 \includegraphics[width=0.45\textwidth] {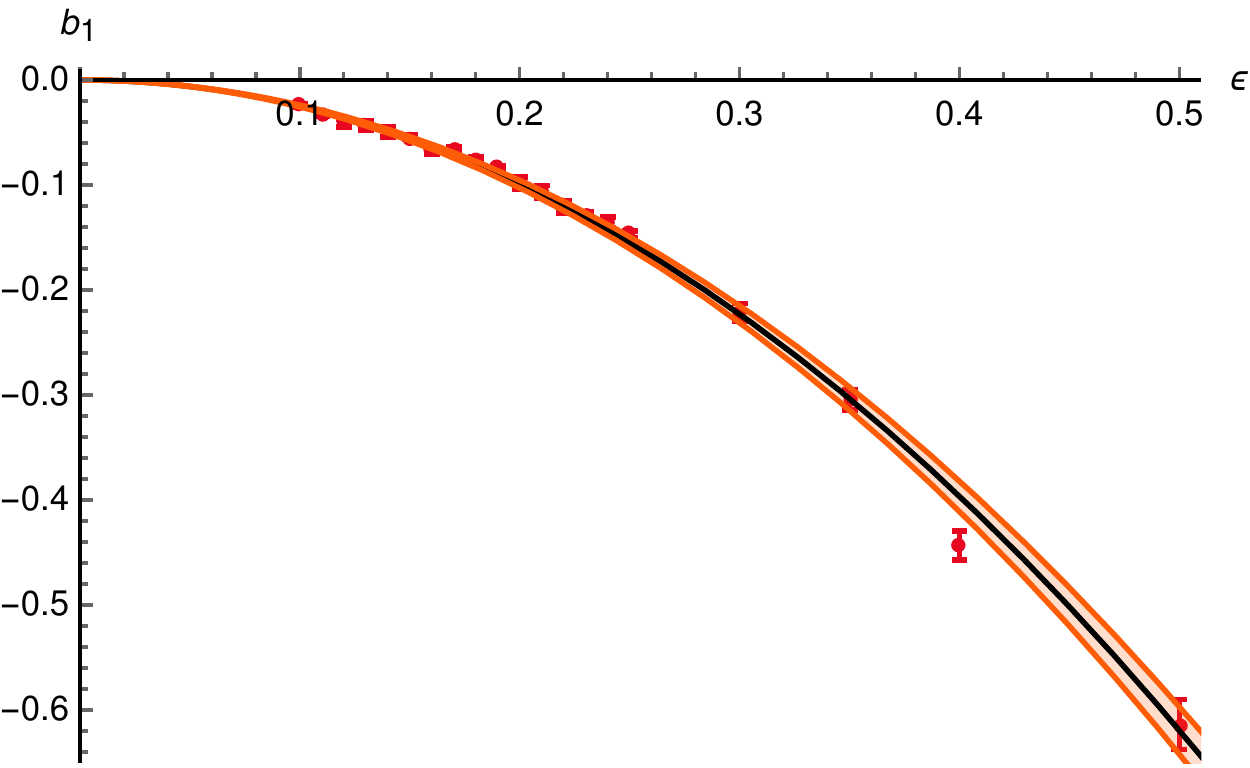}
\end{center}
\caption{\label{APTfitsb.fig}  Best fits for   $b_0$ and $b_1$ as functions of $\epsilon$.}
\end{figure}
\begin{align}\label{eq:afterPTfullfit}
a^+(N,\epsilon)&=-25.21 (\pm 8.16) +1.53 (\pm 0.21) N, \\
b^+(N,\epsilon)&=0.4 (\pm 0.24 )+ 17.63 (\pm 4.76) \epsilon^2 -2.48 (\pm0.09) \epsilon^2 N\;.
\end{align}
We plot Equation (\ref{eq:pastPTfit}) with these fits in Fig
\ref{bSAPTfits.fig} and find them to be in
good agreement with  the data.
 \begin{figure}
\begin{center}
\includegraphics[width=0.75\textwidth]{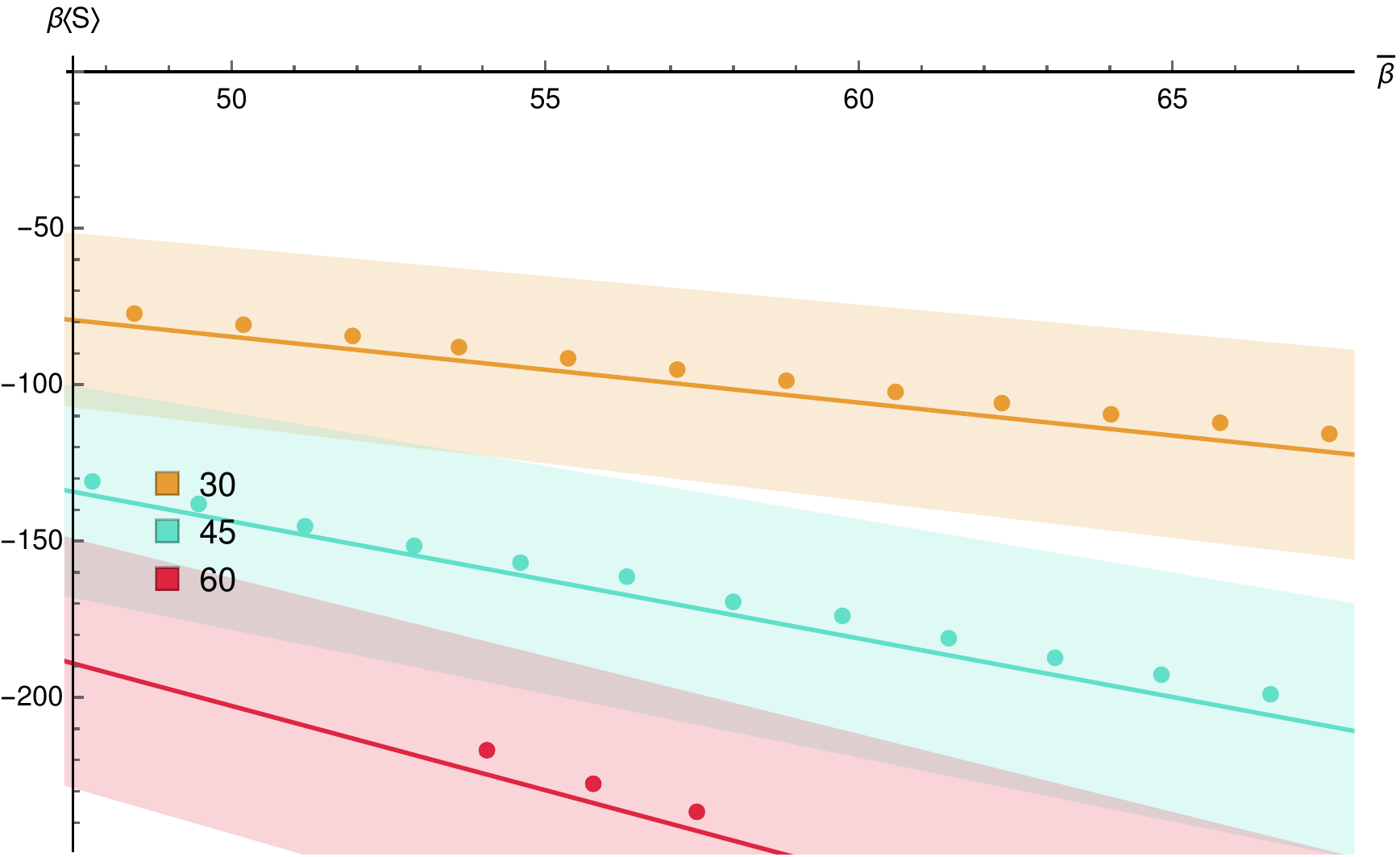}
\end{center}
\caption{\label{bSAPTfits.fig} Plotting the fit using equation
  \eqref{eq:afterPTfullfit} against the data we see good agreement
  with the data, shown here for $\epsilon=0.21$. We plot a few values
  of $N$ to avoid too much overlap in the $99\%$ uncertainty region.}
\end{figure}

We conclude that $\nu_+=1$, which implies that the  free energy $\beta F $ scales as $N$ in this
phase and is hence extensive. Notice that this is consistent with our
expectations in the  large $\beta$ limit, where $S$ for the  bilayer poset scales
as $N^2$.

We note that  this behaviour is reminiscent of the  two phases in
the dimer model of~\cite{dimer} where again $\beta F$ is scale
invariant in one phase and  extensive in the other phase.
\begin{figure}
\centering{
\includegraphics[width=0.75\textwidth]{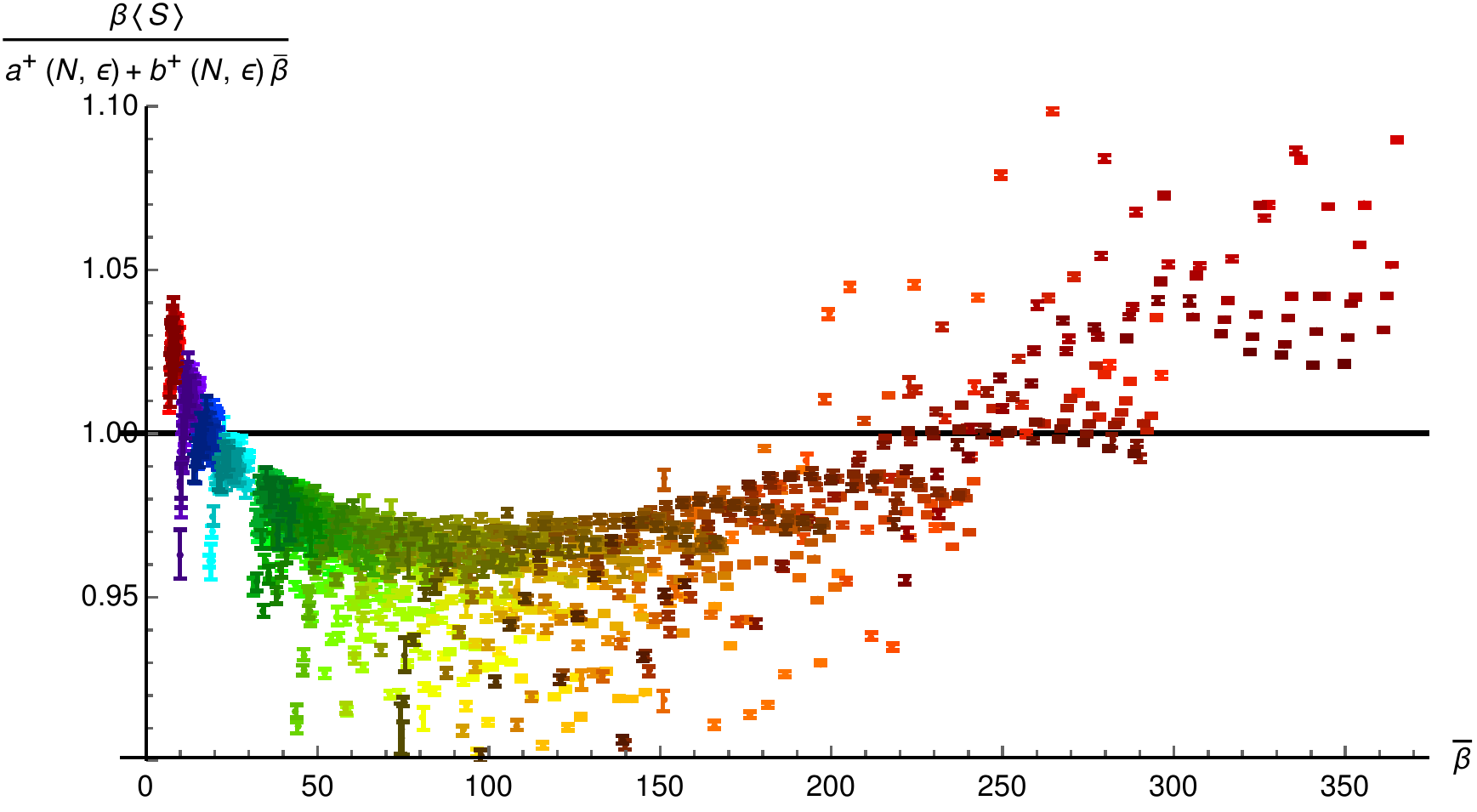}
\caption{Dividing the average action times $\beta$ by the best fit
  function for the phase $\Pi_+$ \eqref{eq:pastPTfit} shows a very
  good fit for all ranges of $N,\epsilon$.
 The color shows the value
  of $\epsilon$ (with $\epsilon$ decreasing from left to right), while
  the brightness indicates $N$, with darker dots corresponding to larger $N$.}
\label{fig:totalcollapse_beforePT}
}
\end{figure}

We now look at the specific heat plots to check the consistency of our
analysis. Given that
\begin{equation}
C=-\beta^2 \frac{\partial \av{S}}{\partial \beta}=-\frac{\bb^2}{N} \frac{\partial \av{S}}{\partial \bb}
\end{equation}
we see that to leading order in $N$
\begin{equation}
\label{SPHTfit}
C_-=-b_1^-(\epsilon) \bb^2, \quad C_+= a_1^+(\epsilon) N
\end{equation}
The unscaled plots in the two regions are given in Figure
\ref{unscaledspht.fig}.
\begin{figure}
\centering{
 \includegraphics[width=0.75\textwidth]{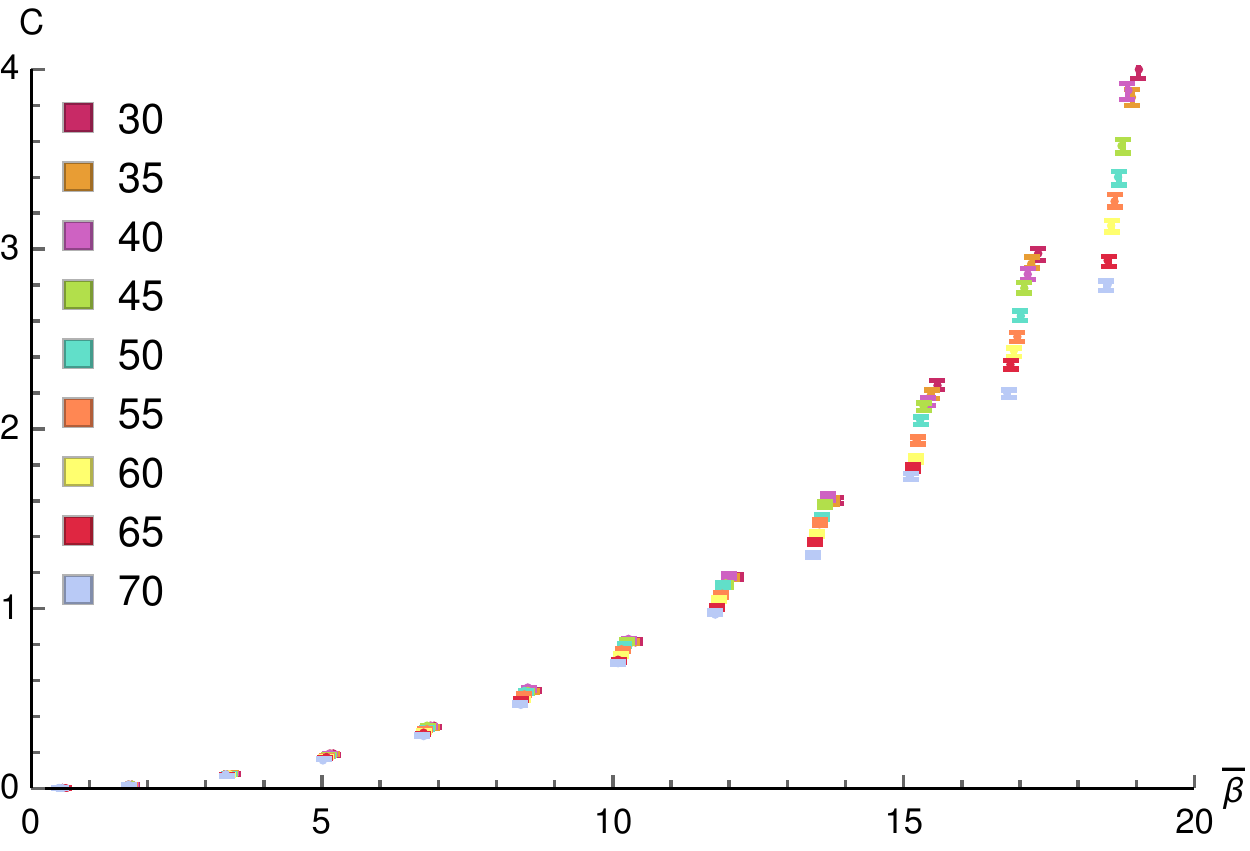}
\vskip 1cm
 \includegraphics[width=0.75\textwidth]{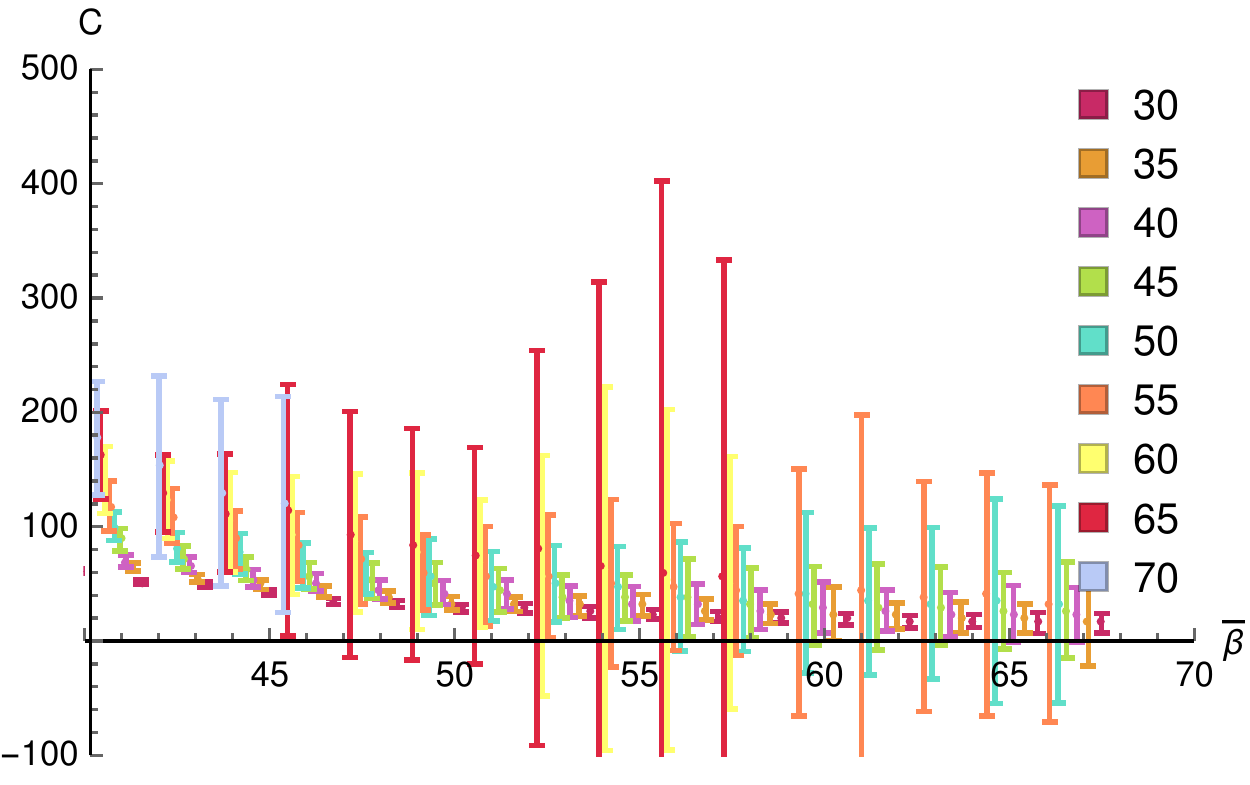}
\caption{\label{unscaledspht.fig} The specific heat $C$ plotted
  against $\bb$ for $\epsilon=0.21$ on either side of the phase transition.}
}
\end{figure}
In the $\Pi_-$ region a collapse is fairly clear, without the need for
scaling. On the other hand,  the $\Pi_+$ region collapses when one scales
by $N$ as shown in Figure \ref{scaledspht.fig}.
\begin{figure}\centering{
   \includegraphics[width=0.75\textwidth]{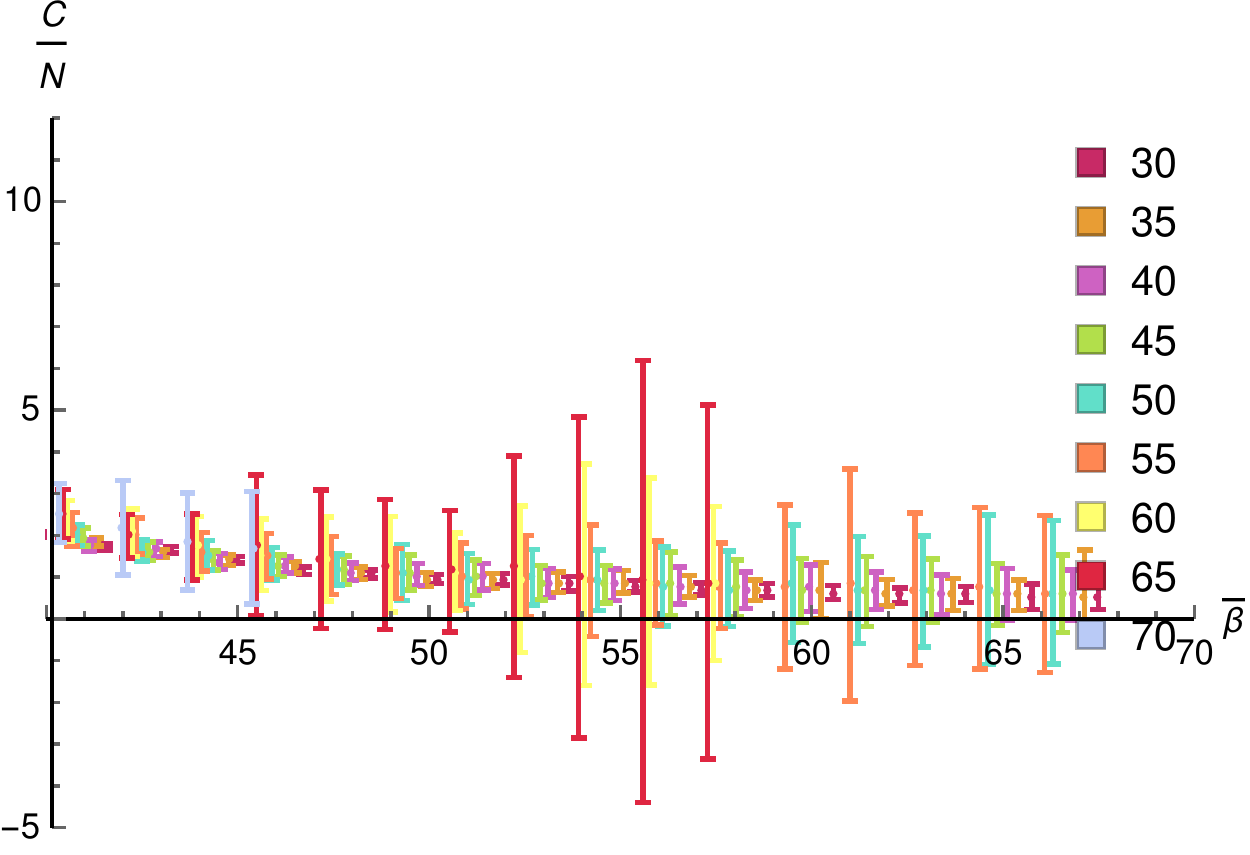}
 \caption{\label{scaledspht.fig} Recaling $C$ by $N$ leads to a very good collapse of the values for all $N$. }}
\end{figure}
The largest  values of $N$ show the  convergence for
the  respective scalings in the two regions.

Plotting the fit lines from eqn \eqref{SPHTfit} doesn not lead to convincing plots.
This is presumably because next to leading order corrections become important further from the phase transition for $C$ than for $\av{S}$.

%


\subsection{First Order Phase Transition}

In Figure \ref{pt.fig} we focus on $\beta \av{S}$ around
the phase transition. We find that as $N$ increases the transition
becomes sharper with $N$ suggesting a discontinuous or first order
transition.

Examining the raw data, we are able to extract the double
Gaussians that characterise first order phase transitions, as shown in
Figure \ref{dg.fig}  at the estimated $\beta_c$ for  $\epsilon=0.21$.
However, as pointed out in \cite{challa}, for finite $N$ these
features alone are not sufficient
to establish conclusively  that the transition is first order.
This is {\it only}  established once we look at the double Gaussians as a function
of $N$. Figure \ref{dg.fig} shows the separation between the
peaks increasing with $N$, which we find is the strongest evidence
for a first order phase transition.

We also analyse the scaling of the peaks of the specific heat $C$.
Since $\nu_+=1 > \nu_-=0$, we expect the scaling at the first order
transition to go as $2 \nu_+=2$.    As shown in Figure
\ref{Cscaling.fig} this is indeed the case.

\begin{figure}
\begin{center}
\includegraphics[width=0.75\textwidth]{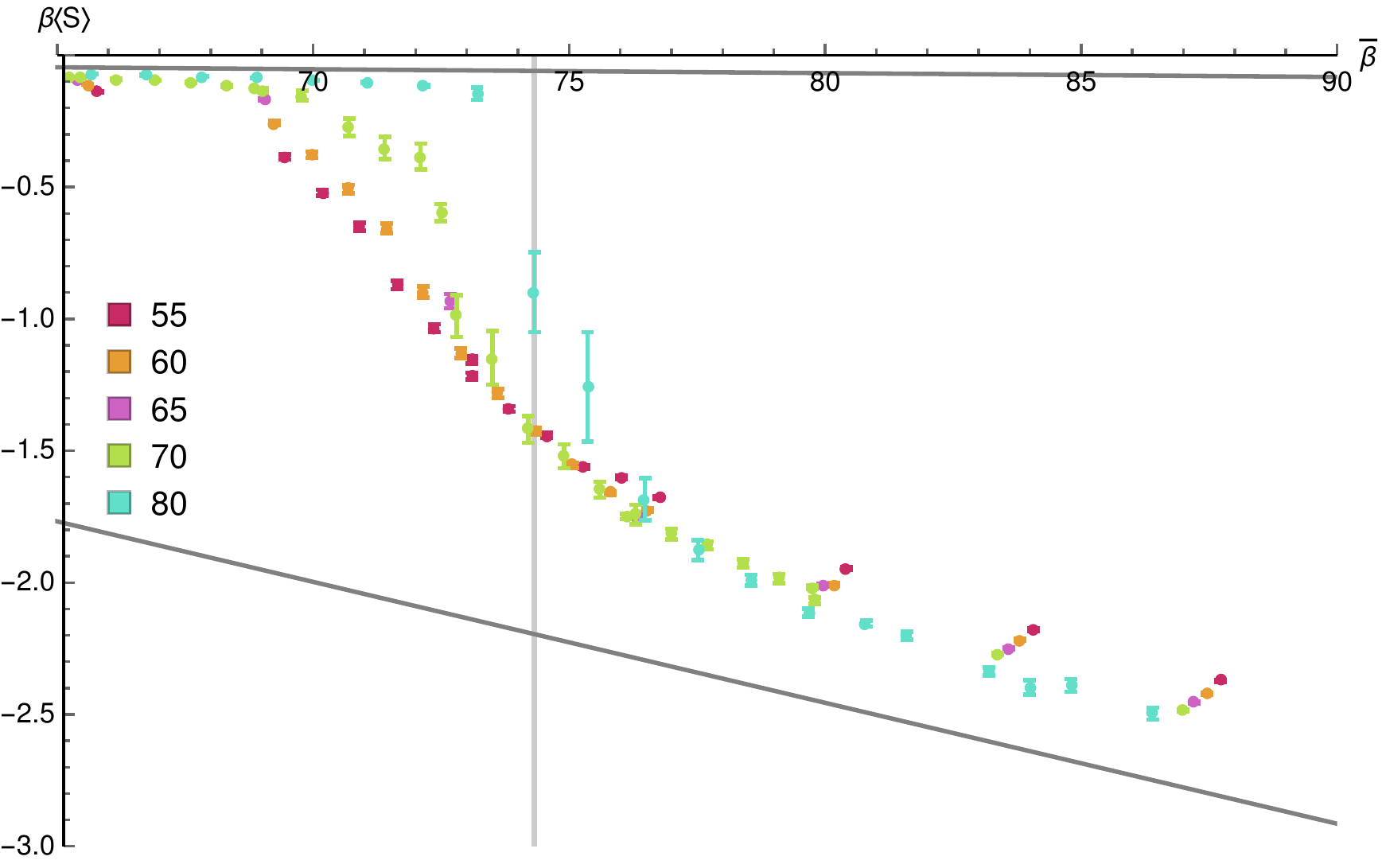}
\caption{\label{pt.fig}
In the region of the phase transition, it is clear that as $N$
increases, $\av{S}$ becomes more and more discontinuous. Here we have
plotted the fit functions for $\av{S}$ before and after the phase
transition to demonstrate the discontinuity.}
\end{center}
\end{figure}

\begin{figure}
\begin{center}
\includegraphics[width=0.75\textwidth]{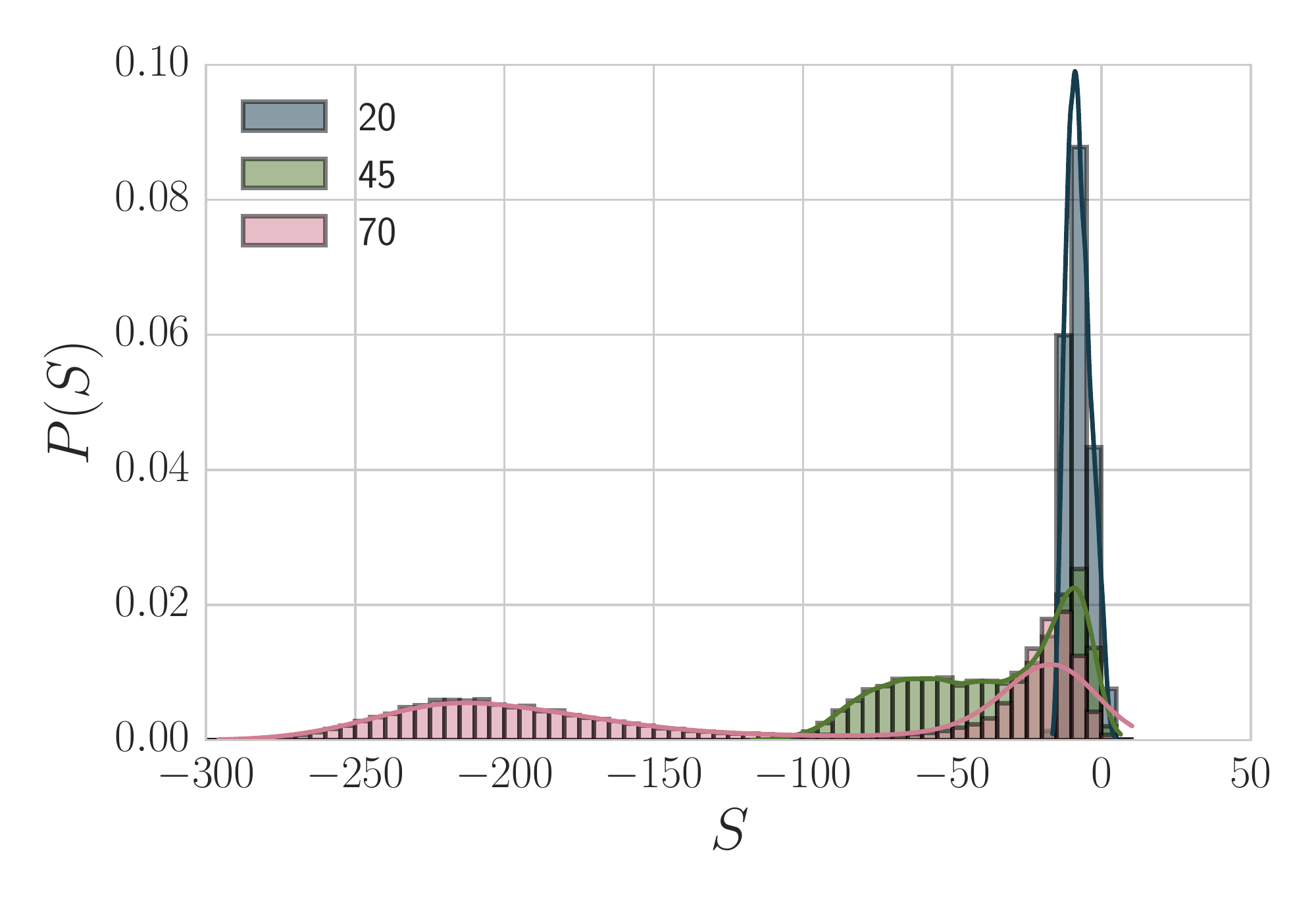}
\caption{\label{dg.fig}
Plotting all measured values of $S$ into a histogram shows that at $\beta_c$ the histogram splits into two peaks which wander further away from each other as $N$ increases.
}
\end{center}
\end{figure}

\begin{figure}
\begin{center}
\includegraphics[width=0.75\textwidth]{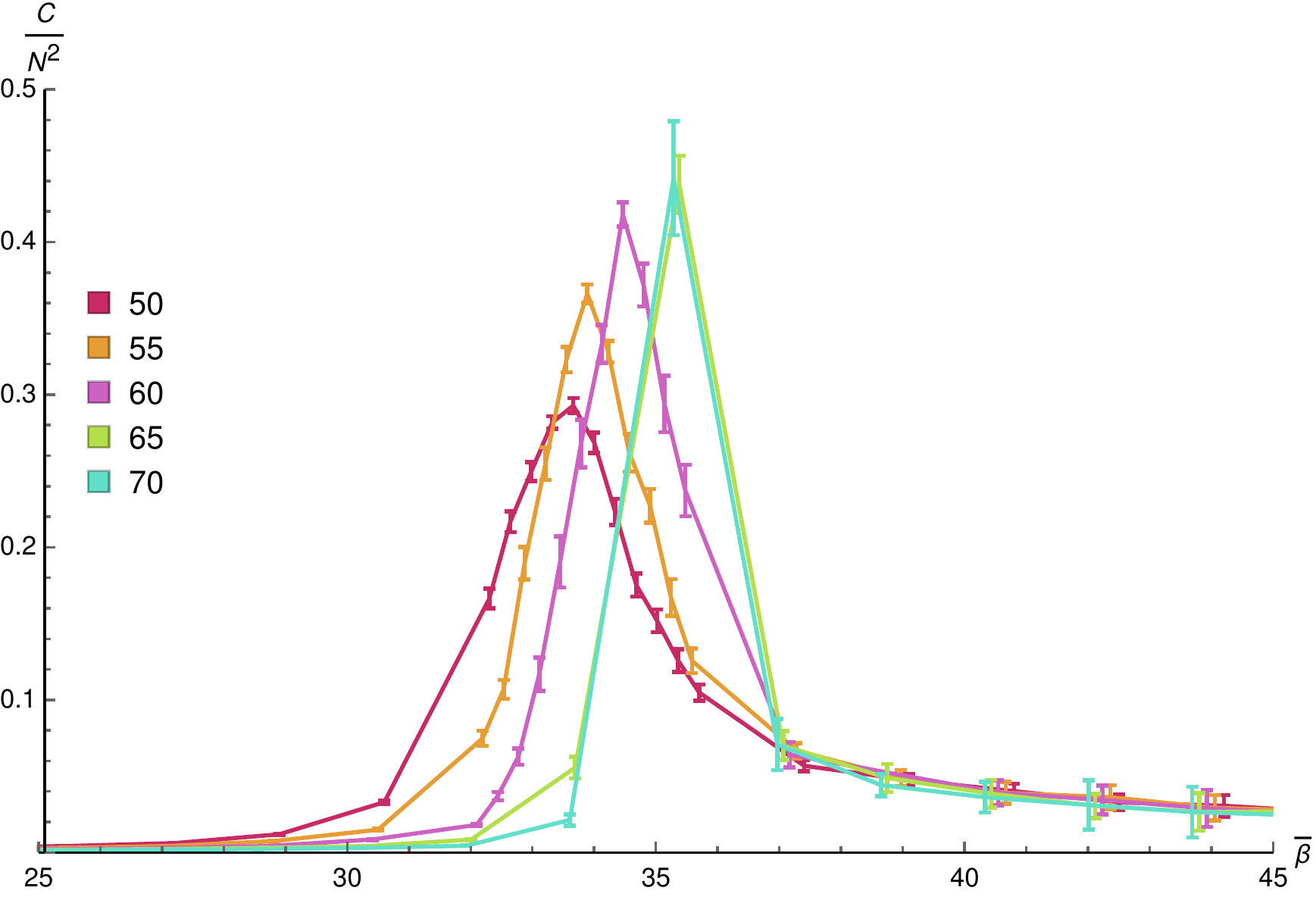}
\caption{\label{Cscaling.fig} We show the rescaling of the peak values
  of $C$ by $N^2$ to converge for larger $N$. Here $\epsilon=0.21$.}
\end{center}
\end{figure}

Another marker to explore the order of a phase transition is the so
called Binder cumulant\footnote{This is shifted by 2/3 from  the
  standard definition.}
\begin{align}
B=\frac{1}{3}\left(1- \frac{\av{S^4}}{\av{S^2}^2}\right)\;.
\end{align}
At fixed $N, \epsilon$ this quantity has a minimum at the
pseudocritical point, and as shown in \cite{challa}
it takes on a non-zero negative value at first order transitions,
and is zero for second order transitions.  While in finite size
systems some deviation from $0$ is expected, this can be determined by
observing the trend as $N$ is increased.

The double Gaussian Eqn (\ref{nong}) predicts for $N \rightarrow
\infty$ and $p_+=t, p_-=1-t$, asumming  that $\mu_+>\mu_-$ that
$B=\frac{1}{3} - \frac{t+(1-t)x^4}{3(t+(1-t)x^2)^2}$, where
  $x\equiv\frac{\mu_-}{\mu_+}$.  When the Gaussians are equally
  weighted ($t=\frac{1}{2}$),  $B$ lies between $0$ and
    $-\frac{1}{3}$}.

\begin{figure}
  \begin{center}
    \includegraphics[width=0.5\textwidth]{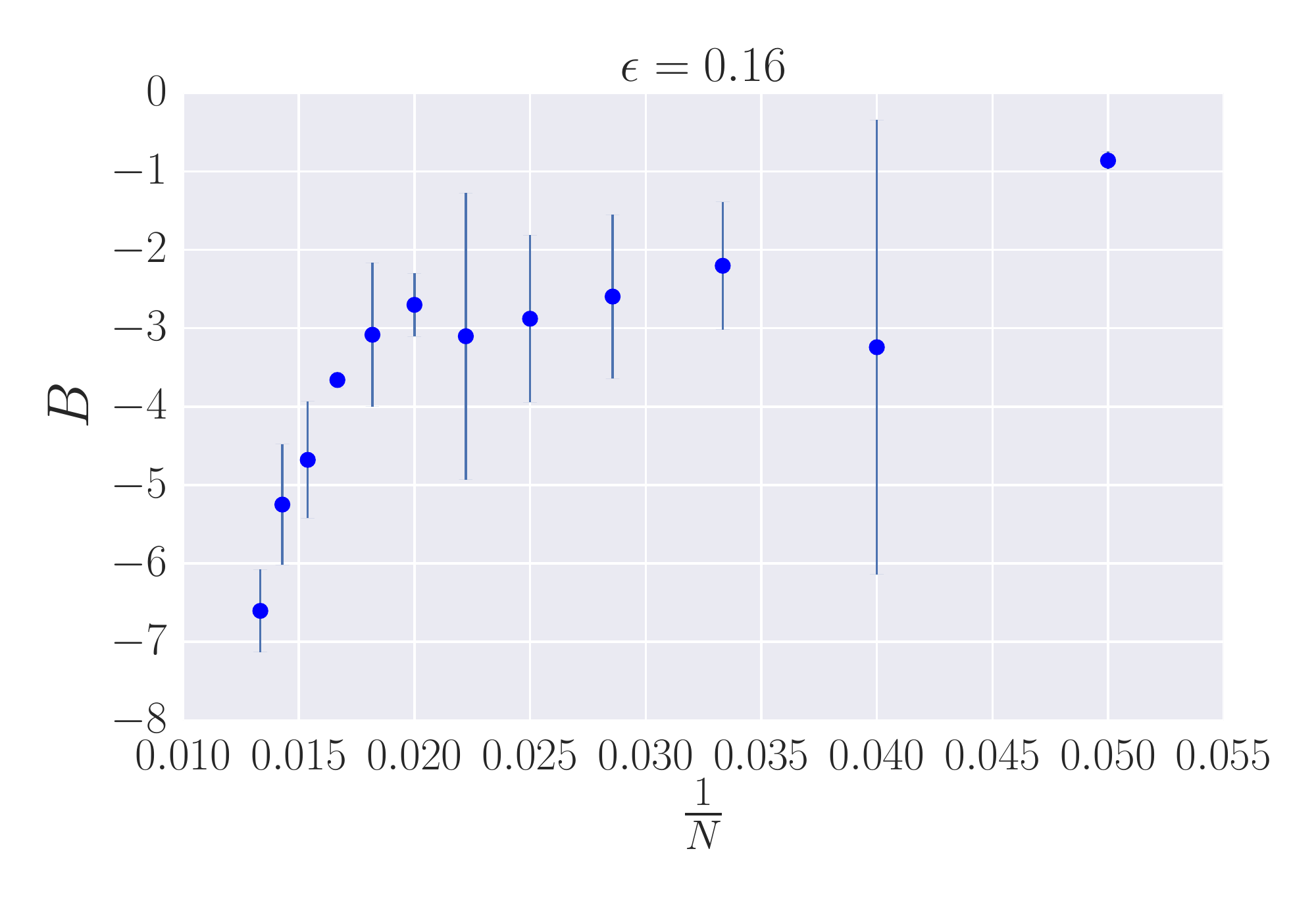}
    \caption{The Binder coefficient moves further away from $0$ as $N$
      becomes larger, thus indicating a first order phase transition.}
    \label{fig.binder}
  \end{center}
 \end{figure}

To be able to determine the minimum value of the Binder coefficient
with reasonable precision,  we used a reweighting procedure, as explained in \cite{newman_monte_1999} to approximate the value of $B$ in the region where we expected the pseudocritical point.
We plot this observable agains the inverse system size in figure
\ref{fig.binder}, and can clearly see that it tends to a non-zero
value.   Taking $\mu_+=\av{S_+} \sim N^2,
  \mu_-=\av{S_-} \sim N$ gives $B \sim \frac{1}{3}-\frac{1}{3t}$  for large
  $N$.  For larger $N$, we see that $t\sim 0.04$, which suggests that we have not sampled the
  immediate vicinity of the critical point with sufficient
  precision. Alternatively it may be due to the asymmetry of scaling
  on either side of the phase transition. In either case, Fig
  \ref{fig.binder} provides additional supporting evidence for a first
  order phase transition.


\section{Conclusions and Open Questions}
\label{five.sec}

The most unexpected and physically  interesting outcome of our
analysis is the possible generation, of a negative non-zero
cosmological constant $\Lambda$ associated with the constant
curvature spacetime $\mathrm{adS_2}$. The scale of $\Lambda$,
moreover,
is simply set by $\epsilon$ and $\bb$, via the relation
\begin{equation}
\Lambda = \kappa_2 b_1^-(\epsilon) \bb,
\end{equation}
where $\kappa_2 \sim l_p^{-2}$. While there is no {\it natural} Planck
scale in $\twod$, given a scale, $\Lambda$ can be made
as  small as desired by choosing $\bb$  to be small enough.

Apart from the extensivity of the action, there are other indications
that  a typical  causal set in this phase is approximated by a causal
patch of $adS_2$.  The  Myrheim-Myer
(MM)  flat spacetime dimension is given by the inverse of the
ordering fraction $f_r=r/{{N}\choose {2}}$ where $r$ denotes the number of
relations \cite{myr,meyer}. In our simulations we have  generated this
ordering fraction alongside the action. We find that  in the continuum phase, our data shows
that $f_2 \sim 0.5$ resulting in an MM flat spacetime dimension of
$\sim 2$, which lends support to this being a continuum phase.

\begin{figure}
\begin{center}
\subfloat[][]{\includegraphics[width=0.35\textwidth]{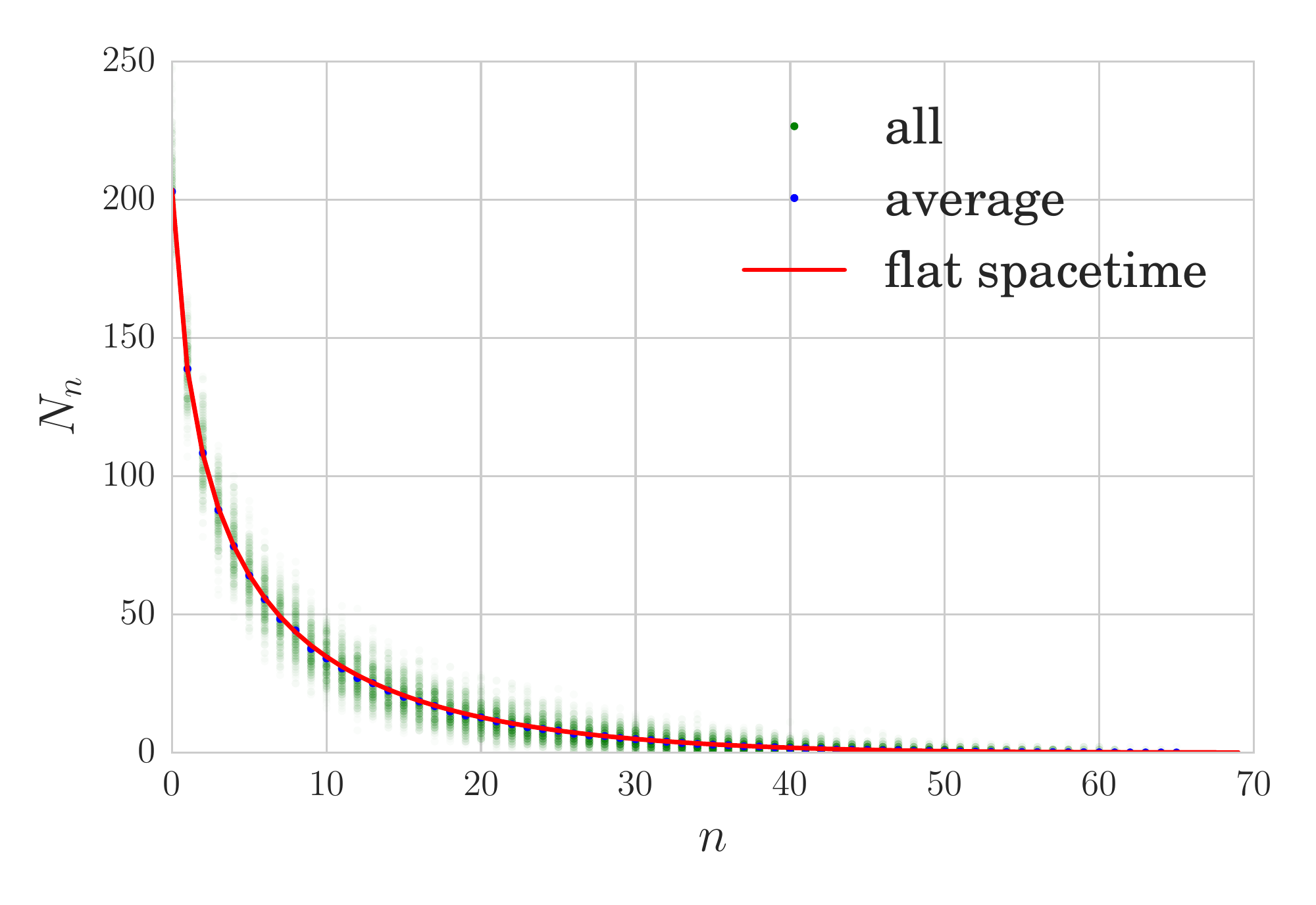}}
\subfloat[][]{\includegraphics[width=0.35\textwidth]{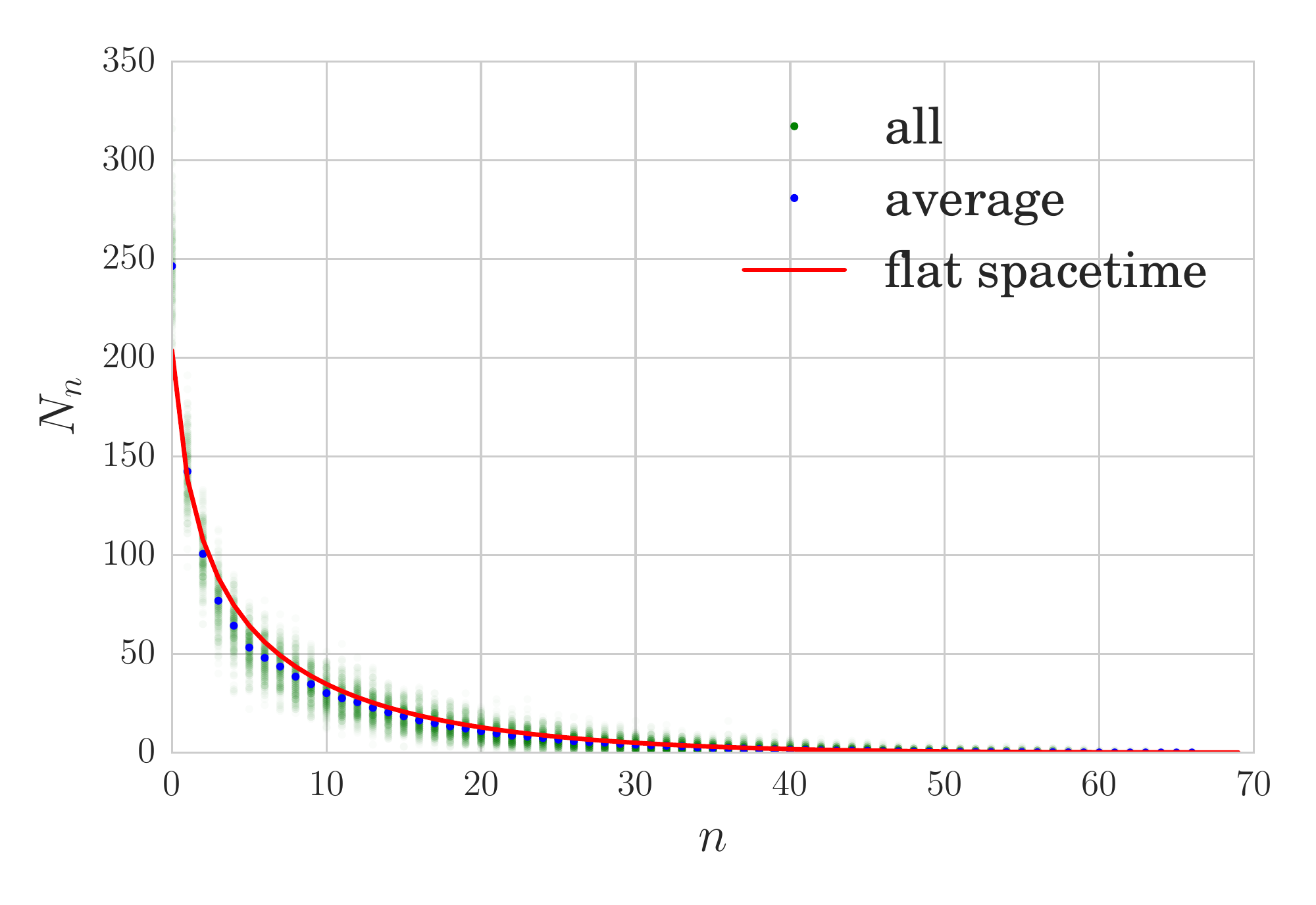}}
\subfloat[][]{\includegraphics[width=0.35\textwidth]{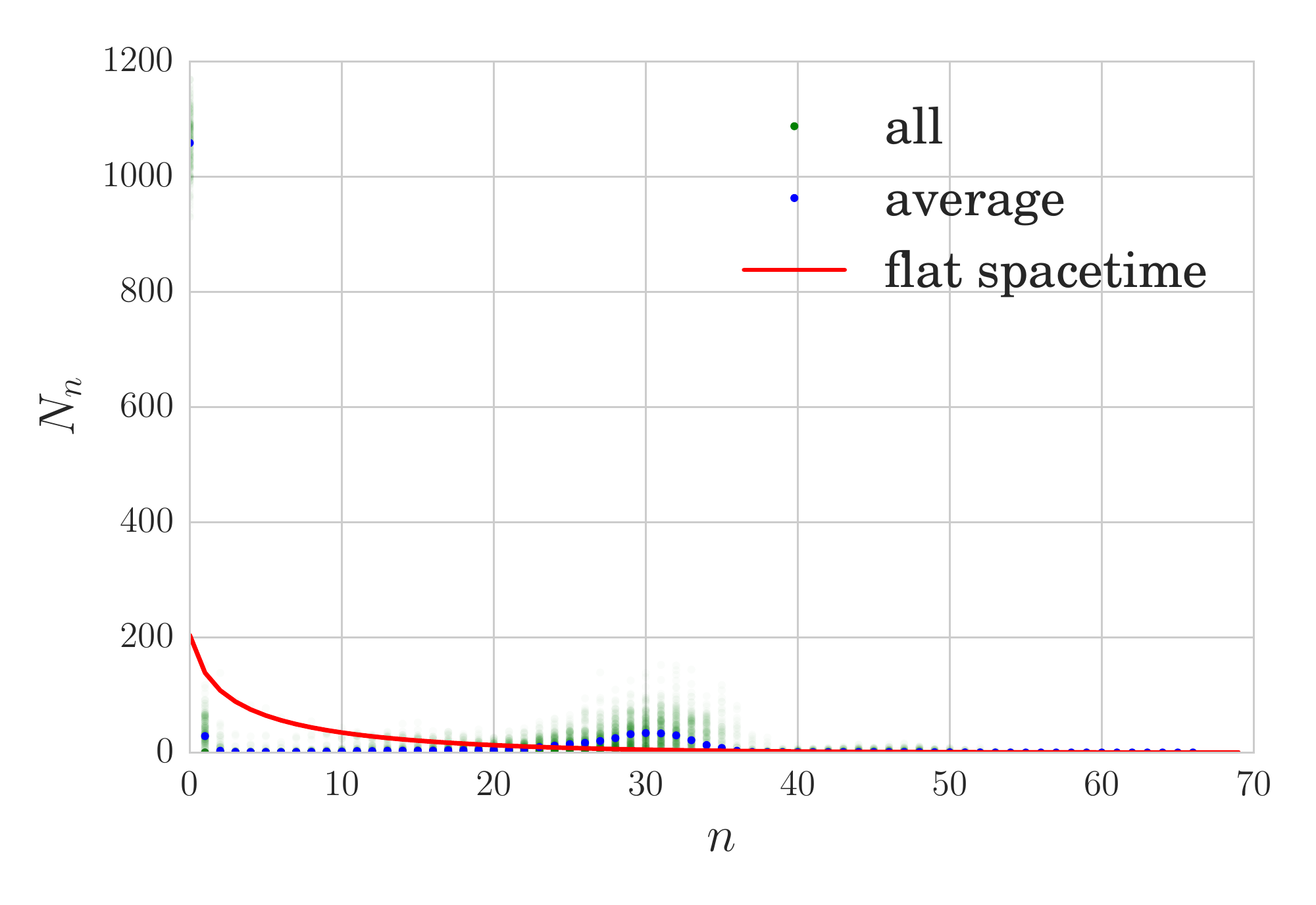}}
\caption{\label{adS.fig} For the smallest values of $\bb$ the
  abundance of intervals is indistinguishable from flat spacetime (a), but
begins to deviate as $\bb$ is increased (b). After the phase
transition(c), the deviation is very pronounced and corresponds to the crystalline phase }
\end{center}
\end{figure}

Additionally, in our simulations we have generated and saved
actual configurations, which allows us to extract other observables of
interest, like the abundance of intervals $\av{N_n}$. We find  that up until the phase transition
they satisfy the expected continuum behaviour (\cite{glaser_towards_2013}) as seen in Fig \ref{adS.fig}(a)-(c).
 For very small $\beta$ in Fig \ref{adS.fig}(a) the
abundance curve simply tracks that of flat spacetime. However, as
$\beta$ increases, $\beta<\beta_c$, the abundance curve dips below
that of flat spacetime as shown in Figure \ref{adS.fig}(b). While we
do not at present have an exact comparisons
 with sprinklings into adS
spacetime, we note that Figure 12 of \cite{glaser_towards_2013} shows
that for positive $\Lambda$, i.e., de Sitter spacetime, the curve
rises above the flat spacetime curve, while the curves of matter and
radiation dominated FRW dip below this curve. Thought of as a postive
pressure, the negative $\Lambda$ therefore seems compatible with the
latter.

In the above analysis we have probed
thermodynamic quantities, which can frequently be reliably obtained
from small systems. This is due to the existence of a limiting free
energy whose finite size corrections fall off with inverse powers of
$N$. Our analysis suggests a clear approach to the asymptotic regime
 by
$N \gtrsim 65$ for $\twod$ CST. Given that we refer to phase
$\Pi_-$ as a ``continuum'' phase, this might seem surprisingly small
to recover continuum behaviour, but this can be traced to the
non-locality in the causal set.  For example, in a simplicial
decomposition or triangulation of $\twod$ spacetime, the valency of
the dual graph is $3$ and so
 the amount of information contained in
an $N$ element simplex grows
 like $3 N \sim O(N)$. It is only by
increasing this (local) {\sl
 information} that one can hope to
find continuum-like behaviour.

On the other hand non-locality implies  that
that the information in a causal set can be very much
 larger
than the $O(N)$ suggested by its cardinality. The information in
 a causal set approximated by flat spacetime is contained in the
 abundance profile $\av{N_n}$ \, v/s  $n$, where $\av{N_n} \sim N \ln N$
in the asymptotic limit for $\twod$ spacetimes, with order $N$
corrections \cite{glaser_towards_2013}. The simultaneous imposition of
this form on each $n$ grows at a {\it minimum} as $N \ln N$. An exact estimation
of this is out of the scope of the present paper.

Finally we note that our discussion of the large $\beta$ limit
generalises to the BD action in any dimension, which has a function
similar to  $f(n,\epsilon)$ with a maximum at
$n=0$. Hence
 configurations with the maximum number of links will
have the smallest
 action and hence the lowest energy.  If we
consider the full sample space of $N$
 element causal sets, there is
an entropic dominance of so-called Kleitman-Rothschild(KR) posets
which have three layers and hence are non-manifold like.  Their
density of states goes as $\sim
 2^{\frac{N^2}{4}}$ for large
$N$~\cite{kr}.  On the other hand, the number of links in each KR
poset $N_0 \sim N^2/8$ with the next significant abundance being
for $n={N}/4$,  whose contribution to the action is highly
suppressed by $f(n,\epsilon)$ for large enough $\epsilon$. Thus the
contribution of the KR posets
$\sim 2^{\frac{N^2}{4}} {\rm e}^{+\beta\epsilon^2 N^2 }$ which for
large enough $\beta$ is subdominant to the contribution from the
bilayer
 posets $ \rho_{b}(\beta, N, \epsilon)
 {\rm e}^{+2\beta\epsilon^2 N^2 }$, Equation (\ref{dom}). Hence we predict that the
large $\beta$ phase will be dominated by the bilayer posets for the
class of statistical partition functions defined by the BD action in
any dimension.

\section{Acknowledgements}
We thank David Rideout for help with Cactus. The simulations were
performed using the Cactus Causal Set Toolkit
\cite{cactusorig,cactus}.  This work has been supported by COST Action MP1405 ``Quantum structure of spacetime (QSPACE).
 LG has received funding from the People
Programme (Marie Curie Actions) H2020 REA grant agreement n.706349
"Renormalisation Group methods for discrete Quantum
Gravity". This research was supported in part by
Perimeter Institute for Theoretical Physics. Research at Perimeter
Institute is supported by the Government of Canada through the
Department of Innovation, Science and Economic Development and by the
Province of Ontario through the Ministry of Research and Innovation.
SS was also supported in part under an agreement with Theiss Research
and funded by a grant from the FQXI Fund on the basis of proposal
FQXi-RFP3-1346 to the Foundational Questions Institute. SS is
currently supported by FQXi-MGA-1510 and an Emmy Noether Fellowship,
at the Perimeter Institute.

\bibliography{orders}
\bibliographystyle{unsrt}
\end{document}